\pdfoutput=1
\newcommand*{\ATLASLATEXPATH}{latex/}
\documentclass[cernpreprint,texlive=2016,txfonts,UKenglish]{latex/atlasdoc} 
\usepackage{\ATLASLATEXPATH atlaspackage}
\usepackage{\ATLASLATEXPATH atlasbiblatex}
\usepackage{\ATLASLATEXPATH atlascontribute}
\usepackage{\ATLASLATEXPATH atlasphysics}
\graphicspath{{logos/}{figures/}}
\usepackage{Tracking2015-defs}
\AtlasTitle{Performance of the ATLAS Transition Radiation Tracker in
Run 1 of the LHC: tracker properties}

\author{The ATLAS Collaboration}

\AtlasJournalRef{JINST 12 (2017) P05002}

\AtlasDOI{10.1088/1748-0221/12/05/P05002}

\AtlasAbstract{ 
The tracking performance parameters of the ATLAS Transition Radiation Tracker (TRT) as part of the ATLAS inner detector are described in this paper
for different data-taking conditions in proton--proton, proton--lead and lead--lead collisions at the Large Hadron Collider (LHC). The performance is studied using data collected
during the first period of LHC operation (Run 1) and is compared with Monte Carlo simulations. The performance of the TRT, operating with two
different gas mixtures (xenon-based and argon-based) and its dependence on the TRT occupancy is presented.
These studies show that the tracking performance of the TRT is similar for the two gas mixtures and that a significant contribution 
to the particle momentum resolution is made by the TRT up to high particle densities. 
}

\hypersetup{pdftitle={Performance of the ATLAS Transition Radiation Tracker in Run 1 of the LHC: tracker properties},pdfauthor={The ATLAS Collaboration}}
\begin{document}

\maketitle
\tableofcontents
\clearpage

\hyphenation{ATLAS}

\section{Introduction}
\label{section:introduction}

The ATLAS detector~\cite{atlas} at the Large Hadron Collider (LHC) is a general-purpose detector designed to make precision measurements of known
physics processes and to search for new physics at the energy frontier of the LHC. At the centre of the detector is an optimised, multi-technology tracking detector immersed in a 2\,T axial magnetic field produced by a solenoid. 
This central tracking detector, also called the inner detector (ID)~\cite{ATLAS:1997ag}, is designed to provide a 
high-precision reconstruction of charged-particle trajectories (``tracks'').
The ID was successfully commissioned in 2008~\cite{Aad:2010bx}, and since 2009 was operated with efficiency greater than 97\% during 
the first period of the LHC operation (Run 1 in 2009--2013).

\begin{figure}
  \centering
  \includegraphics[width=1.0\columnwidth]{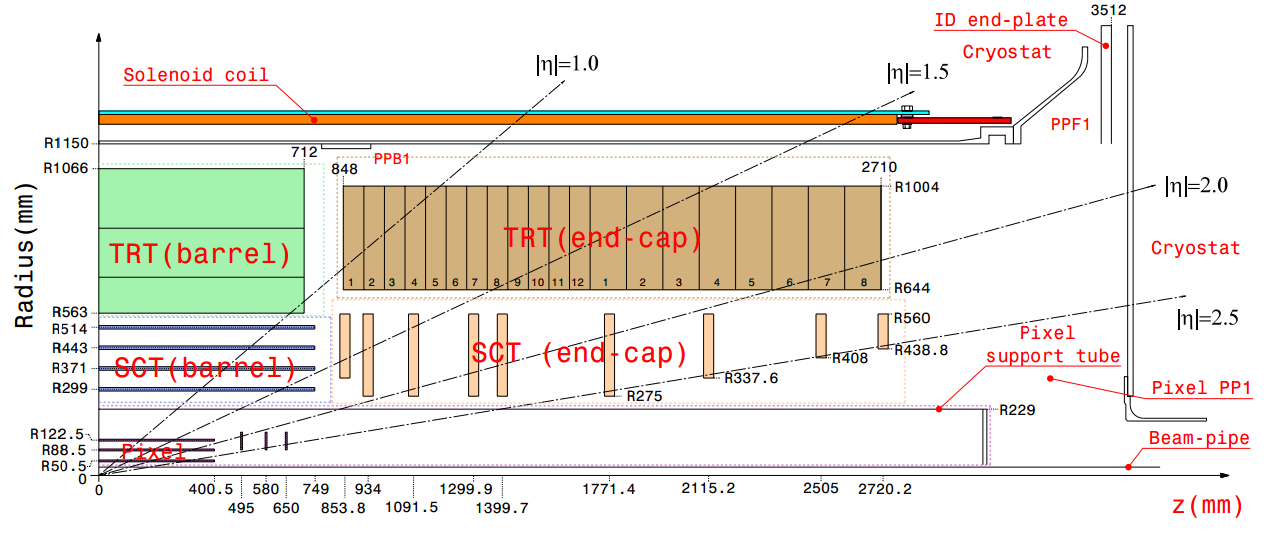}
  \caption{
  The inner detector of the ATLAS experiment (during the first LHC data-taking period in 2009--2013) is viewed in the $r$--$z$ plane, where $r$ is the radial distance from the $z$-axis. The detector is symmetric with respect to the plane perpendicular to the beam line and passing through the collision point (0,0). Only the positive $z$ side is shown.}
 \label{fig:AtlasInnerDetector}
\end{figure}

\begin{figure}[!h]
  \centering
  \includegraphics[width=0.350\columnwidth]{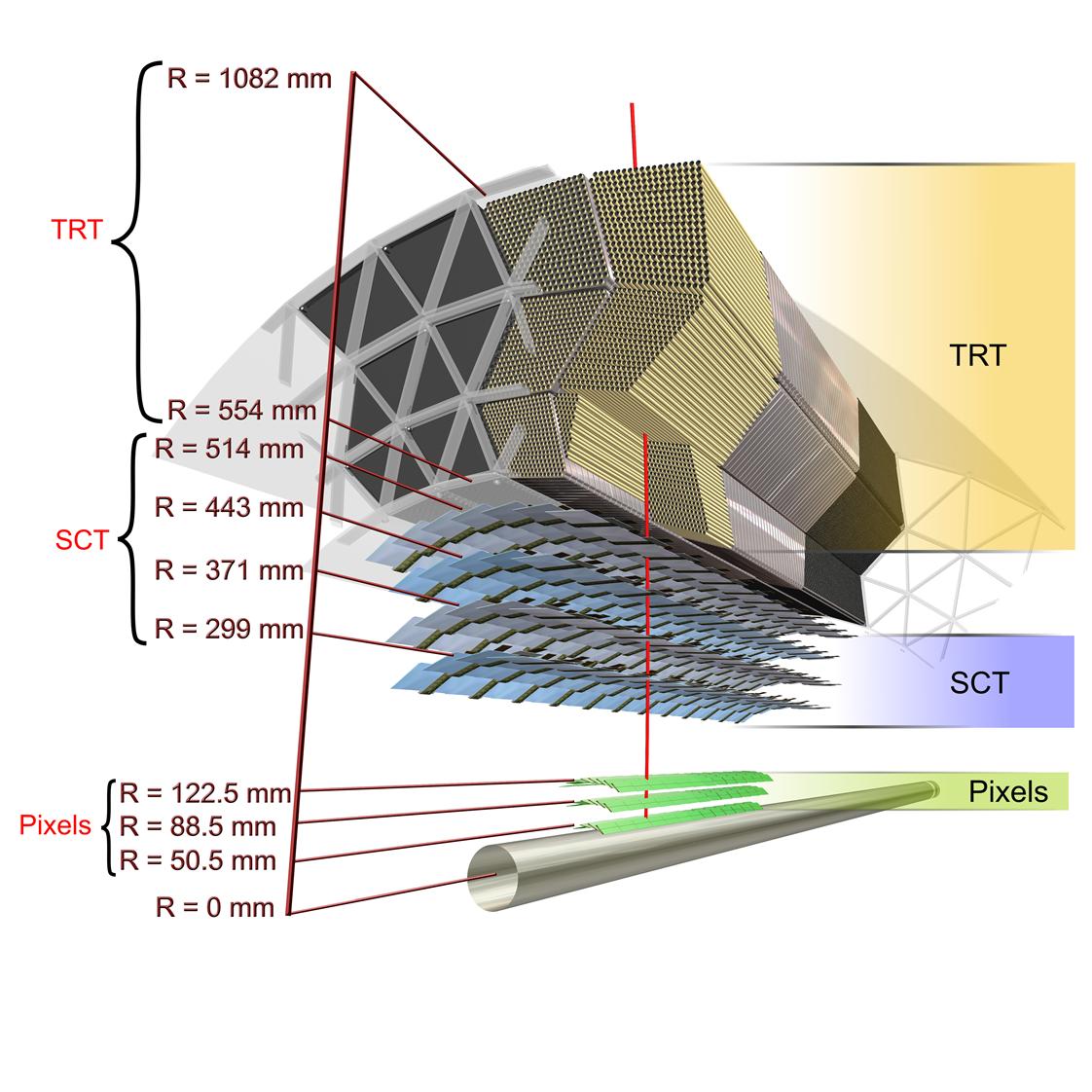}
  \label{fig:TRT_straws_barrel}
  \includegraphics[width=0.550\columnwidth]{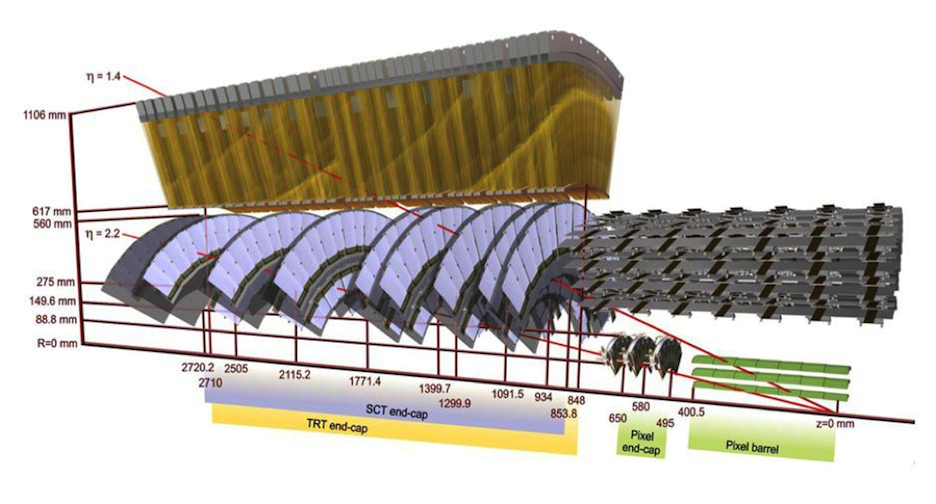}
  \label{fig:TRT_straws_endcap}
\caption{
A schematic view of the ATLAS ID barrel (left) and left end-cap side (right). Particle tracks crossing the ID are shown by red lines. 
}
\label{fig:TRT_straws}
 \end{figure}

The ID covers the pseudorapidity range $|\eta| <2.5$ and has full coverage in $\phi$\footnote{\label{note1}The ATLAS reference system is a Cartesian right-handed coordinate system, with the nominal collision point at the origin. The positive $x$-axis is defined as pointing from the collision point to the centre of the LHC ring and the positive $y$-axis points upwards, the $z$-axis is defined along the beam pipe. The azimuthal angle, $\phi$ is measured around the beam axis, and the polar angle, $\theta$ is measured with respect to the $z$-axis. The pseudorapidity is defined as $\eta=-\ln(\tan(\theta /2))$.} (see Figure~\ref{fig:AtlasInnerDetector}).
It consists of a silicon pixel detector at the innermost radii (pixel), surrounded by a  silicon microstrip detector (SCT), and a straw-tube detector 
called the Transition Radiation Tracker (TRT) 
that combines continuous tracking capabilities with particle identification based on transition radiation (TR). Each detector consists of a barrel and 
two end-caps (see Figure~\ref{fig:TRT_straws}), where the positive (negative) $z$ side is named ``A'' (``C'').

The precise measurement of the particle trajectories is fundamental for most of the data analyses in ATLAS.
The positions of signals in individual detector elements (``hits'') are used to reconstruct tracks inside the tracker. 
The track reconstruction algorithms employ local and global pattern recognition to identify a coherent pattern of hits corresponding to a track produced by a charged particle. 
An iterative track-fitting procedure is used to estimate the compatibility of the detector measurements with the track hypothesis and to compute the final 
track parameters \cite{Aad:2009wy}. 
These measurements correspond to the combination of 
neighbouring hits called ``clusters'' in the pixel and SCT detectors and to the distance of closest approach of the charged-particle track to each straw in the TRT.  

\begin{figure}
  \centering
  \includegraphics[width=0.6\columnwidth]{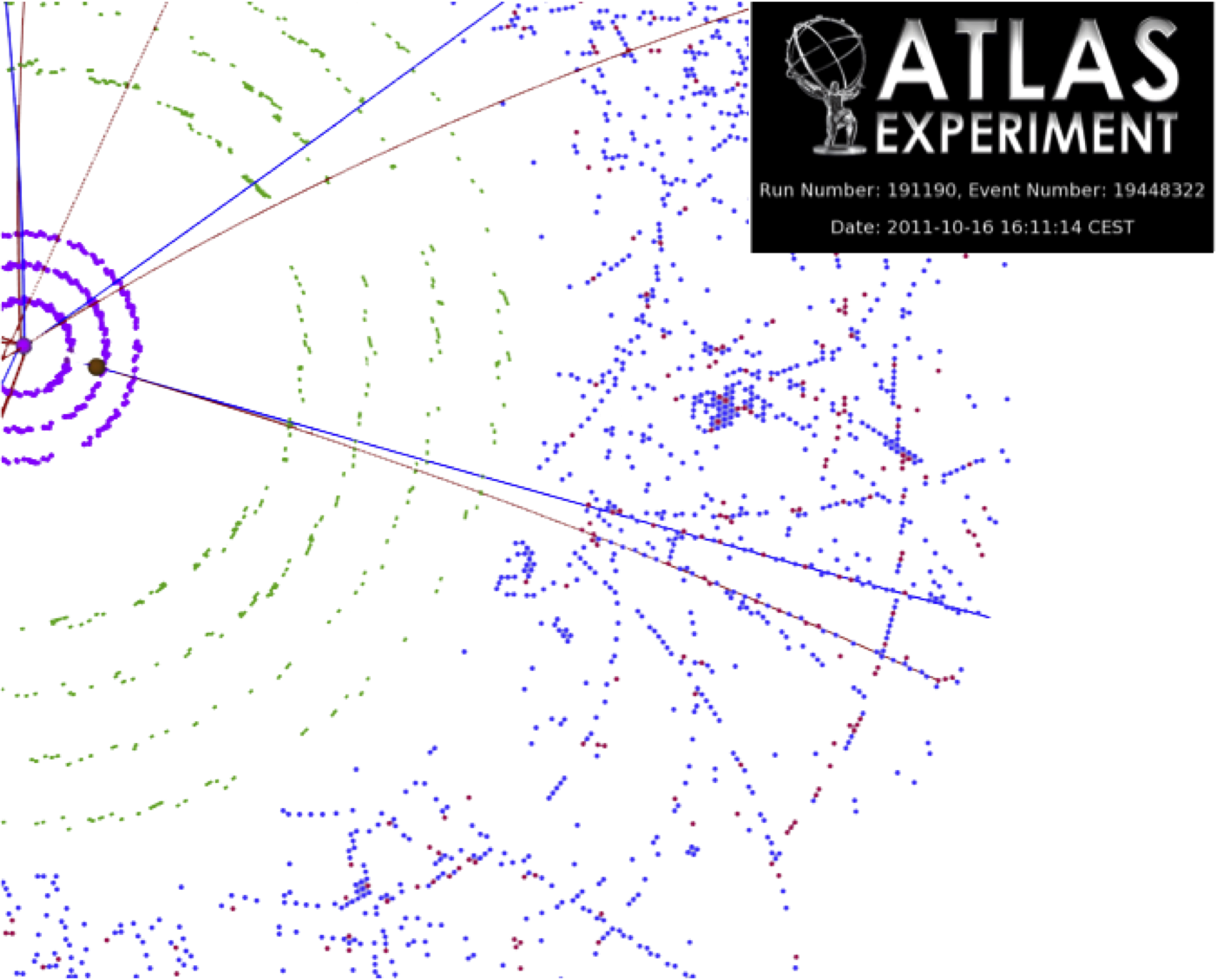}
  \caption{ 
An event display containing a close-up view perpendicular to the beam direction of hits and reconstructed tracks in the ID with $p_{\mathrm{T}}>2~\GeV$  and $|\eta|<1.4$.
Only the hits in the pixel, SCT and TRT layers within $-1<\eta<0$ are shown. 
Here pixel hits ($45.5 < r < 242$ mm) are shown as magenta dots, and SCT space points ($255 < r < 549$ mm) are shown in green. TRT hits ($554 < r < 1082$ mm) above tracking threshold 
are shown as blue dots, with red dots corresponding to hits above the TR threshold. A photon conversion vertex (large brown dot in the second pixel layer) as well as the associated  reconstructed electron (blue line) and positron  (red line) tracks are also shown.}
\label{fig:trt_event_display}
\end{figure}

Figure ~\ref{fig:trt_event_display} shows an event display illustrating reconstructed tracks in the ID with $p_{\mathrm{T}} > 2~\GeV$ and $|\eta|<1.4$.
A track from a particle traversing the ID volume with $|\eta|<2$  would typically have 3 pixel clusters, 8 SCT strip clusters, and more than 30 TRT straw hits.

Another important feature of the ATLAS ID is the particle identification capability implemented in the TRT.
The particle identification is based on TR, which arises when ultra-relativistic charged particles cross a boundary between 
media with different dielectric constants~\cite{Artru:1975xp}. The probability of a given particle to emit transition radiation is determined solely by its Lorentz $\gamma$-factor. 
In the X-ray energy range of the TR photons, this probability is a step-like function which rises 
from about zero to its maximum when the $\gamma$-factor changes from about 500 to 2000. 
The TRT was built to exploit this effect, in particular to distinguish between electrons and pions up to momentum values of $\sim 100~\GeV$.

During Run 1 the TRT operated with an efficiency greater than 97\%, contributing significantly to the particle 
momentum measurement, pattern recognition accuracy and particle identification.

The TRT tracking performance for different running conditions in Run 1, which include three types of collisions (proton--proton, proton--lead, and lead--lead collisions) and imply 
different TRT straw occupancies, are presented in this paper.
The straw occupancy characterises the particle density in the TRT and is defined as the probability to have a straw signal above a given threshold value in the TRT read-out window of 75 ns, equivalent to three bunch crossings at the nominal 25 ns LHC bunch spacing.
For proton--proton ($pp$) collisions, the occupancy depends on the average number of interactions per bunch crossing $\langle\mu\rangle$.
The $pp$ collisions studied in this paper are collected at $\sqrt{s}=8~\TeV$ either with a 50 ns bunch interval in low straw-occupancy conditions 
or with individual isolated bunches in high straw-occupancy conditions. Lead--lead collisions (called later in the text ``heavy ion'' collisions) are collected 
at $\sqrt{s}=2.76~\TeV$ with a 200 ns bunch interval and proton--lead collisions are collected at $\sqrt{s}=5.02~\TeV$ with a 50 ns interval.  

The simulations used to describe the data are reported in Section~\ref{section:simulation}.
A general description of the TRT, including a brief summary of the data acquisition system is given in Section~\ref{section:trt}.
Position coordinates of hits in the TRT are measured using drift-time information. The straw coordinate accuracy depends strongly on the calibration of the $r$--$t$ relation, where $r$ is the track position radius in the straw and $t$ is the measured drift time. Section~\ref{section:calib} describes the $r$--$t$ calibration procedure.
The track reconstruction accuracy substantially depends on precisely knowing the position of all TRT straw anode wires.
A dedicated TRT alignment procedure was developed and the results of this procedure are presented in Section~\ref{section:TRT_alignment}.
Results from studies of the TRT tracking parameters at low straw-occupancy are presented in Section~\ref{section:xenon} for straw tubes filled with a xenon-based (Xe-based) gas mixture, which was the baseline operation during Run 1. The TRT tracking parameters are the straw efficiency, the straw track position measurement accuracy, and the number of TRT precision hits. These parameters are defined later in the text. 

During the 2012 data-taking period, several leaks developed in the gas pipes which bring the active gas to the cleaning and mixing stations. In most cases, the leaks are located in inaccessible areas and their repair is not possible. Because of the high cost of losing the Xe-based gas mixture, the possibility to operate the most affected modules with a significantly less expensive argon-based (Ar-based) gas mixture was investigated. In order to understand the TRT performance with such an Ar-based mixture, dedicated studies were performed during the proton--lead collisions in 2013 where leaking modules in the barrel and end-caps were supplied with the Ar-based mixture. The results of these studies are presented in Section~\ref{section:argon}. TRT modules with high leak rates have been routinely operated with the Ar-based gas mixture since the beginning of the second period of LHC operation, Run 2, which started in 2015.

At an LHC design luminosity of 10$^{34}$ cm$^{-2}$s$^{-1}$ a TRT straws occupancy can reach 60\%. Large detector 
occupancies present many challenges for track reconstruction, including a degradation of the track parameter resolution due 
to incorrect hit assignments, a decrease of hit efficiencies, and an increase in the rate of fake tracks due to random hit combinations.
Studies of the TRT tracking capabilities in a high-density particle environment are reported in Section~\ref{section:occupancy}. For these studies,  special LHC 
high-intensity $pp$ collision fills with $\langle\mu\rangle$ up to 70 and heavy-ion collision runs were analysed.

In Section~\ref{section:jets}, tracking performance studies in jet cores at different TRT occupancies are presented.
Conclusions are presented in Section~\ref{section:conclusion}.

\section{Simulation}
\label{section:simulation}

Simulated events are produced with various Monte Carlo event generators, processed through a full ATLAS detector 
simulation~\cite{SOFT-2010-01} using \textsc{Geant4}~\cite{bib-geant4}, and then reconstructed with the same procedure as for data.

For the  studies with a Xe-based gas mixture, $Z\rightarrow \mu\mu$ events, with a $\langle\mu\rangle$ distribution between 5 and 30, are simulated  using 
the $\SHERPA$ 1.4 generator \cite{Gleisberg:2008ta} with the CT10 parton distribution function (PDF) set~\cite{Lai:2010vv}
and leading-order (LO) matrix elements with up to five additional final-state partons.
For the studies with an Ar-based gas mixture, samples with single muons are used. Muons are generated with $\eta$ and $p_{\mathrm{T}}$ 
distributions similar to the ones observed in proton--lead collisions in 2013 data.

For the high-occupancy studies, a special pile-up-only sample with a uniform $\langle\mu\rangle$ distribution between 38 and 71 is used. The same conditions 
for bunch spacing as in data are simulated. 
The response of the ATLAS detector and the performance of the reconstruction algorithms for heavy-ion collisions are evaluated using simulated minimum-bias 
events produced with the version 1.38b of the HIJING event generator~\cite{Wang:1991hta}, and using the same bunch-spacing and pile-up conditions as in data.

Performance in a dense track environment is studied using simulated $pp$ dijet events from the 
$\PYTHIA$ 8.1~\cite{Sjostrand:2007gs} generator with parameter values set to the AU2 CT10 tune~\cite{ATLAS:2012uec}.

\section{The Transition Radiation Tracker}
\label{section:trt}

\subsection{General description}
The TRT is a straw tracker composed of 298,304 carbon-fibre-reinforced kapton straws, 4\,mm in diameter, with 70 $\mu m$ walls and held at a potential of $-1530$~V with respect 
to a $31\, \mu m$ diameter gold-plated tungsten ground wire at the centre~\cite{Abat:2008zza}. 
The TRT has two different geometrical arrangements of straws. The barrel section, where 52,544 straws are aligned parallel to the direction of the
beam axis, covers the radial region $560<r<1080$\,mm and the longitudinal region $|z|<712$\,mm~\cite{Abat:2008zzb}. Two end-cap sections, 
each with 122,880 straws that are aligned perpendicular to the beam axis and point outwards in the radial direction, cover the 
regions $644<r<1004$ \,mm and $827<|z|<2744$\,mm~\cite{Abat:2008zz}. The TRT acceptance range is $|\eta| <2.0$.

The TRT barrel consists of 73 layers of straws organised in 3 cylindrical rings (green colour in Figure~\ref{fig:AtlasInnerDetector}), 
each containing 32 identical and independent modules.
All the straws are embedded in stacks of polyproplylene or polyethylene fibres which produce transition radiation X-rays used for electron identification. 
The distance between straws is on average 6.8 mm in both the radial and azimuthal directions, and the length of the straws is 142.4 cm. 
In order to reduce the straw occupancy, the anode wires are separated electrically by  
glass wire joints at $z=0$  so that each side is read out separately.
The straws closest to the beam line have the largest occupancy. To further reduce occupancy, the wires in the innermost 9 of the 73 straw layers 
of the barrel are split into three parts, with only 34\,cm of straw length being read out from each side, while the middle wire is not read out. 
These straws with shorter active region are referred to as short straws. 

Both end-cap sections consist of two sets of identical and independent wheels, in turn organised in sub-wheels (brown colour in Figure~\ref{fig:AtlasInnerDetector}).
The first type of wheels (type A: 12 wheels per side at $827<|z|<1684$\,mm) contains 12288 radial straws positioned in 8 consecutive 
layers spaced by 8 mm along $z$. Each layer contains 768 straws in $r$--$\phi$ and the 4\,mm gap between successive layers is filled with 
radiator material made from thin polypropylene foils. The distance between straws in the  $r$--$\phi$ direction is varied from about 5.2\,mm at 
the innermost radius to 8.4\,mm at the outer radius. The second type of wheels (type B: 8 wheels per side at $1687<|z|<2744$\,mm) 
contains 6144 straws positioned exactly as in the type A wheels in $r$--$\phi$ but with a spacing of 16\,mm in the $z$ direction. 
The space of 12\,mm between straw layers is filled with the same type of radiator material.
The total number of layers in each of the two end-caps is 160 with a straw active length of about 36 cm.

Transition radiation photons (soft X-rays) emitted in the radiator are typically absorbed
by the gas inside the straw tube. During Run 1, all straws were filled with a gas mixture of 70\% Xe, 27\% $\mathrm{CO_2}$, and 3\% $\mathrm{O_2}$. 
The xenon is used for its high efficiency to absorb the TR photons, which have typical energies of 6--15~$\keV$. 
Oxygen is used in this mixture in order to maximise the difference between the operating voltage and the breakdown voltage. 
During Run 2, straws belonging to modules with 
large gas leaks are filled with a gas mixture of 70\% Ar, 27\% $\mathrm{CO_2}$ and 3\% $\mathrm{O_2}$. 
Argon has a much lower efficiency to absorb the TR photons in this energy range, but has similar tracking capabilities as xenon, as it is shown in 
Section~\ref{section:argon}. 

The TRT operates as a drift chamber: when a charged particle traverses a straw, it deposits about 2.5~$\keV$ of energy in the active gas,
 creating about 5--6 primary ionisation clusters per mm of path length. The electrons then drift towards the wire
 and cascade in the strong electric field very close to the wire with a coefficient of $2.5 \times 10^4$ (gas gain), producing a detectable signal. 
The signal on each wire is then amplified, shaped, and discriminated against two adjustable thresholds~\cite{1748-0221-3-06-P06007}: low level (LL) and high level (HL). 
The LL threshold is used to measure an electron drift time for the tracking and is usually set to about 250--300~$\eV$ in absolute scale for the Xe-based gas mixture. 
The HL threshold is used to identify large energy deposits. The separation of particle types is based on the probability
of a particle's signal in a straw to exceed the HL threshold, which is different for electrons which produce TR, and other particles with Lorentz factors
below 1000 which do not produce TR. The HL threshold is optimised to obtain the best electron--pion separation and is set at about 6--7~$\keV$ 
for a Xe-based gas mixture. The settings for an Ar-based gas mixture is discussed in Section \ref{section:argon}.

\subsection{The TRT readout electronics}
The TRT read-out electronics play a crucial role in the detector performance.
A detailed description of the TRT electronics is presented in Ref.~\cite{1748-0221-3-06-P06007}. Here, a basic description
of the front-end electronics design relevant for the understanding of the TRT performance is given.

The TRT front-end read-out system consists of analogue and digital components.
The TRT analogue read-out is based on a custom-designed analogue chip (ASDBLR) which performs amplification, shaping, discrimination, and baseline restoration. 
The peaking time of the analogue signal from the point-like ionisation in the straw gas is about 7.5\,ns,
with a semi-Gaussian shape after ion tail cancellation. The signals after discrimination are sampled by a
second digital chip, the drift-time measuring read-out chip (DTMROC), which performs the
time measurement of the signals and provides a digitised result to the off-detector electronics.

The ASDBLR chip has two discrimination paths which allow the setting of both the LL and HL thresholds.
For LL threshold signals, the DTMROC divides each 25\,ns LHC bunch period into eight equal-time bins so that the LL discriminator
output pulse is recorded in 3.125\,ns bins and stored in a pipeline to that precision. 
For HL threshold signals the information is recorded in the DTMROC chip as a status bit for each of the three 25\,ns intervals inside the TRT read-out window.
In total 24 LL bits and 3 HL bits, representing the information over 3 bunch crossings in the nominal LHC running conditions (75\,ns) are stored in the pipeline for each read-out channel.
Each DTMROC provides timing, storage, and control for two ASDBLR chips.
After the ATLAS Level-1 trigger (which receives and buffers the data from the read-out electronics) accepts an event,
the DTMROC data is shipped to the back-end electronics for further processing and integration.

The ASDBLR chip employs a fixed-time shaping technique in combination with an active baseline restoration
to remove the unwanted signal component from slowly moving ions in the straw gas. 
After this operation, only a small fraction (5\%) of the total avalanche signal is available, corresponding to 0.2 femtoCoulomb (fC) of collected 
charge per primary electron with the TRT gas gain of $2.5 \times 10^4$.

Since the goal is to detect the earliest clusters of electrons arriving at the wire, 
the LL threshold is set to a minimum value defined by a 2\% noise occupancy in each channel,
which is equivalent to a 300\,kHz total noise rate for each channel. The resulting LL threshold is quite uniform across the detector and corresponds to 250-300~$\eV$ in absolute scale ($\sim 2\,\mathrm{fC}$), which corresponds to the energy of 3-4 primary ionization clusters.

The thresholds are kept constant over an entire operation period using regular electronics calibration procedures.

\section{Calibration of the drift-time measurement}
\label{section:calib}

\begin{figure}
  \centering
  \includegraphics[width=0.450\columnwidth]{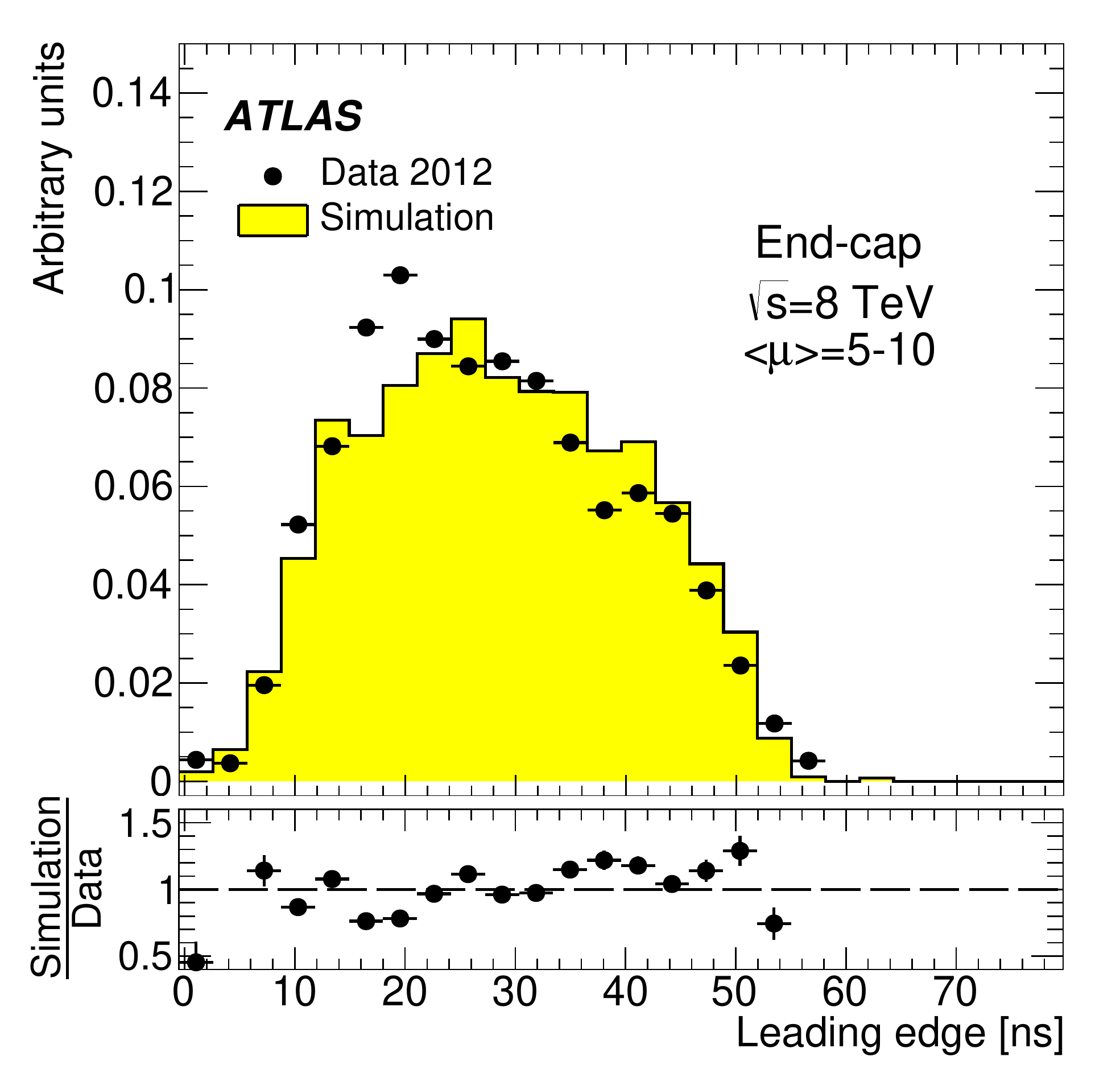}
  \caption{
Distribution of the leading-edge time for hits on track in the TRT end-caps for $5\le\langle\mu\rangle\le10$.
  Data (closed circles) and simulation (histogram) are shown for the 2012 data-taking period operating with the Xe-based mixture. The bins in this histogram correspond to the TRT digitisation binning of 3.125\,ns.} 
 \label{fig:le_EC_t0_3p0_mu_5_10}
 \end{figure}

The spatial position of charged-particle tracks in the TRT detector is determined from the drift-time measurement.
The TRT records the time at which the signal produced by ionisation electrons in the region of closest approach of the track 
to the wire first exceeds the LL threshold, referred to as the leading-edge time. 
It is measured as the time interval between the 40\,MHz LHC clock tick, 
corresponding to a triggered bunch, and the leading edge of the discriminator signal within the TRT read-out window of 75\,ns. An example 
of the leading-edge time distribution for the TRT end-caps straws, comparing data and simulation, is shown in Figure~\ref{fig:le_EC_t0_3p0_mu_5_10}.
\footnote{\label{note2}In all the figures in this paper, $pp$ collisions and the Xe-based gas mixture is used unless another data or mixture types are indicated in the legend.}

The differences between the shapes of the leading-edge distributions observed in data and predicted in simulation are explained by coherent (in time) electronic noise associated
with the 40 MHz clock, which is continuously supplied to the digital part of the TRT front-end electronics. 
The noise modulates the amplifier baseline (ground), changing the effective low-threshold discriminator setting in a way that is
not modelled by the simulation. For signals coming at around 25 ns (see Figure~\ref{fig:le_EC_t0_3p0_mu_5_10}) 
the clock signal raises the baseline, effectively reducing the rise time to reach the low threshold.
This results in an increase of the counting rate and a reduction of the measured drift time. This effect has
an opposite direction for times around 38 ns, where the effective threshold is
increased, reducing the counting rate and increasing the measured drift time.

\begin{figure}
  \centering
  \includegraphics[width=0.5\columnwidth]{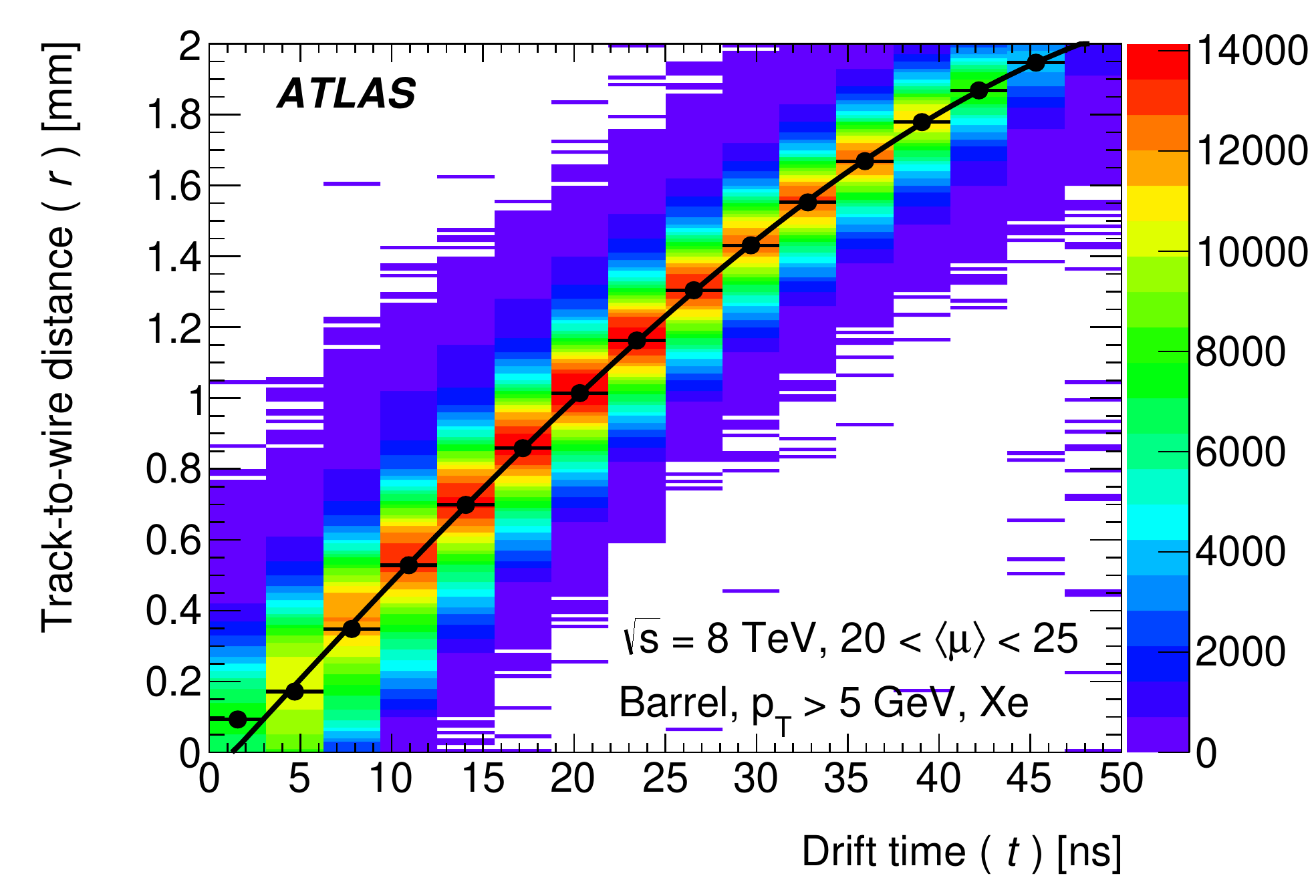} 
  \caption{
The $r$--$t$ relation for the TRT barrel operating with the Xe-based gas mixture during the 2012 data-taking period. The points show the peak position of the fit to the track-to-wire distance ($r$) distribution in 3.125 ns slices of measured drift time ($t$). The fitted line through these points is the relation
used to determine the drift distance based on the measured drift time.}  
\label{fig:r-t}
 \end{figure}

The leading-edge time of the signal in the time window depends on various factors such as the signal propagation in the electronic circuits, 
the time of flight of the particle through the straw and the effect of different threshold settings.
This leads to different shifts in the leading-edge time distributions for straws, which are compensated for by using a parameter $T_{0}$, adjusted during the calibration procedure for each DTMROC chip.
The corrected drift time is obtained by subtracting the value of this parameter to the measured leading-edge time for each chip. 
Once the drift time is known, it can be translated into a drift-circle radius, making use of the relation between drift time and drift distance, called the
$r$--$t$ relation. This relation is obtained from data using the measured drift time and the actual drift radius calculated as the distance of closest approach of the reconstructed track to the anode wire (the so-called ``track-to-wire distance'').
Figure~\ref{fig:r-t} depicts a typical $r$--$t$ relation in the TRT barrel in 2012 $pp$ data. 
The $r$--$t$ relation is determined by fitting the $r$--$t$ data distribution with a third-order polynomial. 
The drift radius is subsequently used by the track fitter during the track-reconstruction procedure.
In the calibration procedure, the $r$--$t$ relation is fitted for each new data-taking session for the four TRT regions (barrel sides A and C, end-cap sides A and C).

To determine the calibration constants ($T_0$ and $r$--$t$ relation) an iterative calibration method has been developed~\cite{alonso:2011}.
Starting from the initial conditions, which correspond to our best knowledge of the detector parameters at a given time, the widths of the time- and position-residuals 
distributions for all hits on tracks are used as figures of merit to determine the break condition in the iterative method. The position residual is 
defined as the difference between the measured drift radius and the track-to-wire distance. The time residual is defined as the difference between the 
measured drift time and the so-called track drift time, defined using the inverse $r$--$t$ relation. 
The calibration constants are re-evaluated using these two figures of merit until the improvement in the width of the residual distributions becomes smaller  
than $\sim 1\%$ (which needs typically two steps).
The new $T_0$ and $r$--$t$ values are used to perform the full reconstruction of the tracks. 

\begin{figure}[!h]
  \centering
  \includegraphics[width=0.450\columnwidth]{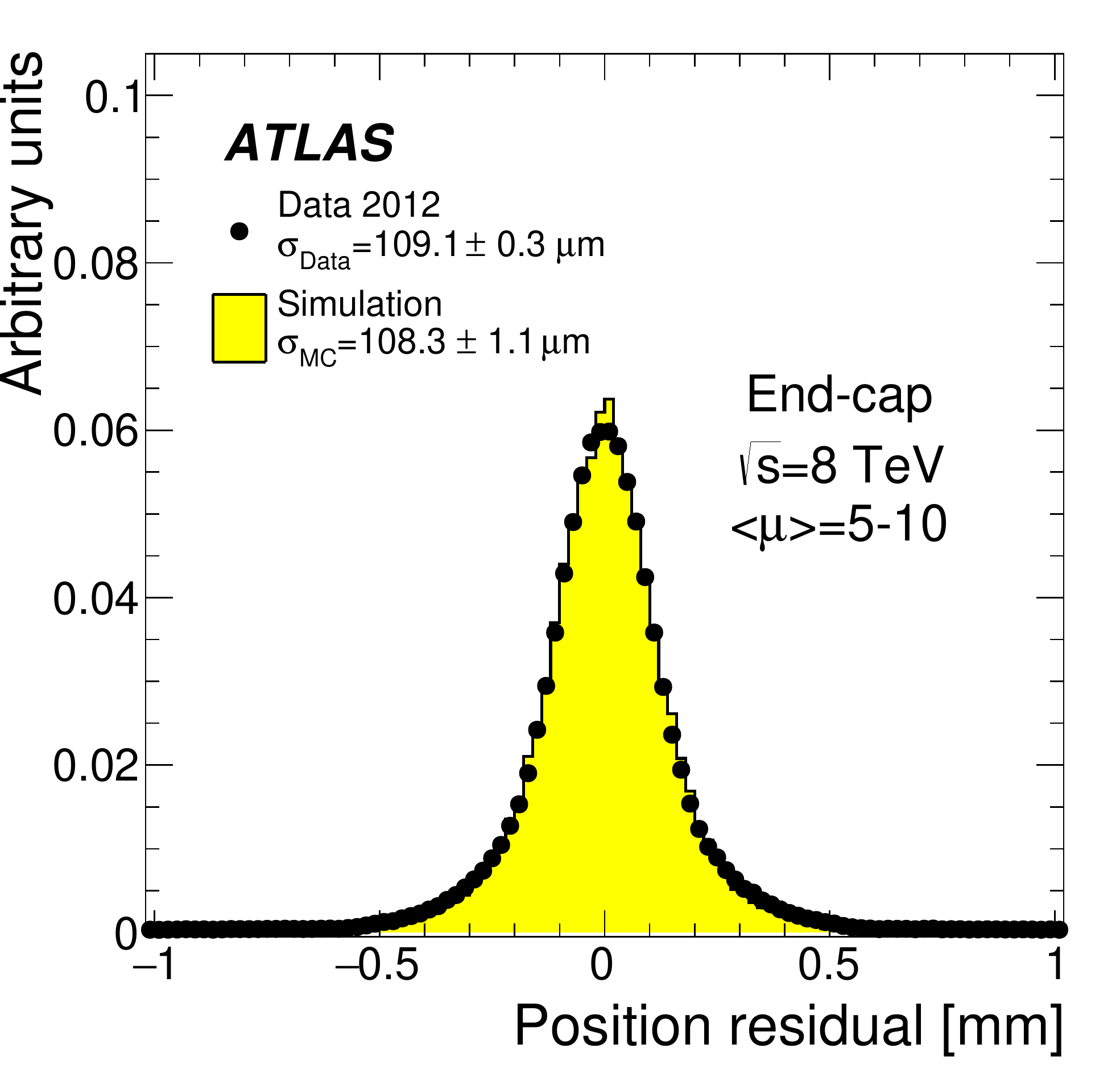}
  \caption{
  Position residual distribution in the TRT end-caps operating with Xe-based gas mixture for muons with $\pt>30~\GeV$.
  Data (solid circles) and simulation (histogram) are shown for a 2012 data-taking period with $5\le\langle\mu\rangle\le10$.
The straw track position measurement accuracy, obtained from an iterative fit procedure to the central $1.5\sigma$ 
of the distribution as described in the text, is  $109.1 \pm 0.3\, \mu$m for the data and  $108.3 \pm 1.1\, \mu$m for the simulation.
  }
  \label{fig:hit_res_unbiased_ec_mu_5_10-ref}
 \end{figure}

An example of the position residual distribution in the end-cap region after the calibration procedure is performed, is shown in 
 Figure~\ref{fig:hit_res_unbiased_ec_mu_5_10-ref}, for $5\le\langle\mu\rangle\le10$. 
The straw track position measurement accuracy is defined in an iterative procedure as the width $\sigma_{\mathrm{f}}$ of a Gaussian function fitted to the peak of the residual distribution within 
the range $\pm 1.5\sigma_i$, where $\sigma_i$ is obtained from the previous iteration. The fitting procedure is repeated  until stable results are achieved.
The straw track position measurement accuracy is about 110 $\mu \mathrm{m}$ in both the barrel and end-caps for $5\le\langle\mu\rangle\le10$.
Good agreement between data and simulation is observed. 
These resolution values are below the typical 130 $\mu$m resolution obtained in test-beam campaigns~\cite{Abat:2008zzb}. 

The uncertainty in the drift-radius measurement is also needed as an input for the track-fitting algorithm.
This uncertainty is defined as the width $\sigma$ of a Gaussian function fitted to the full range of the position residual distribution and thus is expected to be larger than the straw track position measurement accuracy.
This uncertainty (referred to as the ``drift-circle error'') depends on the shortest distance between the anode wire and the particle crossing the straw. 
It is calibrated for both the data and Monte Carlo samples as a function of the drift time.
An accurate determination of this uncertainty is necessary for track reconstruction~\cite{Aad:2009wy}, as it plays a role in the decision whether or not to include 
a given drift-radius  measurement in the track-fitting procedure. 
Once a measurement is added to the track fitting, the uncertainty is also used to determine the weight of its contribution to the track fit. 
TRT hits with a track-to-wire distance within 2.5 times the drift-circle error (referred to as ``precision hits'') have the largest weight in the track fit.

\begin{figure}[!h]
  \centering
  \includegraphics[width=0.450\columnwidth]{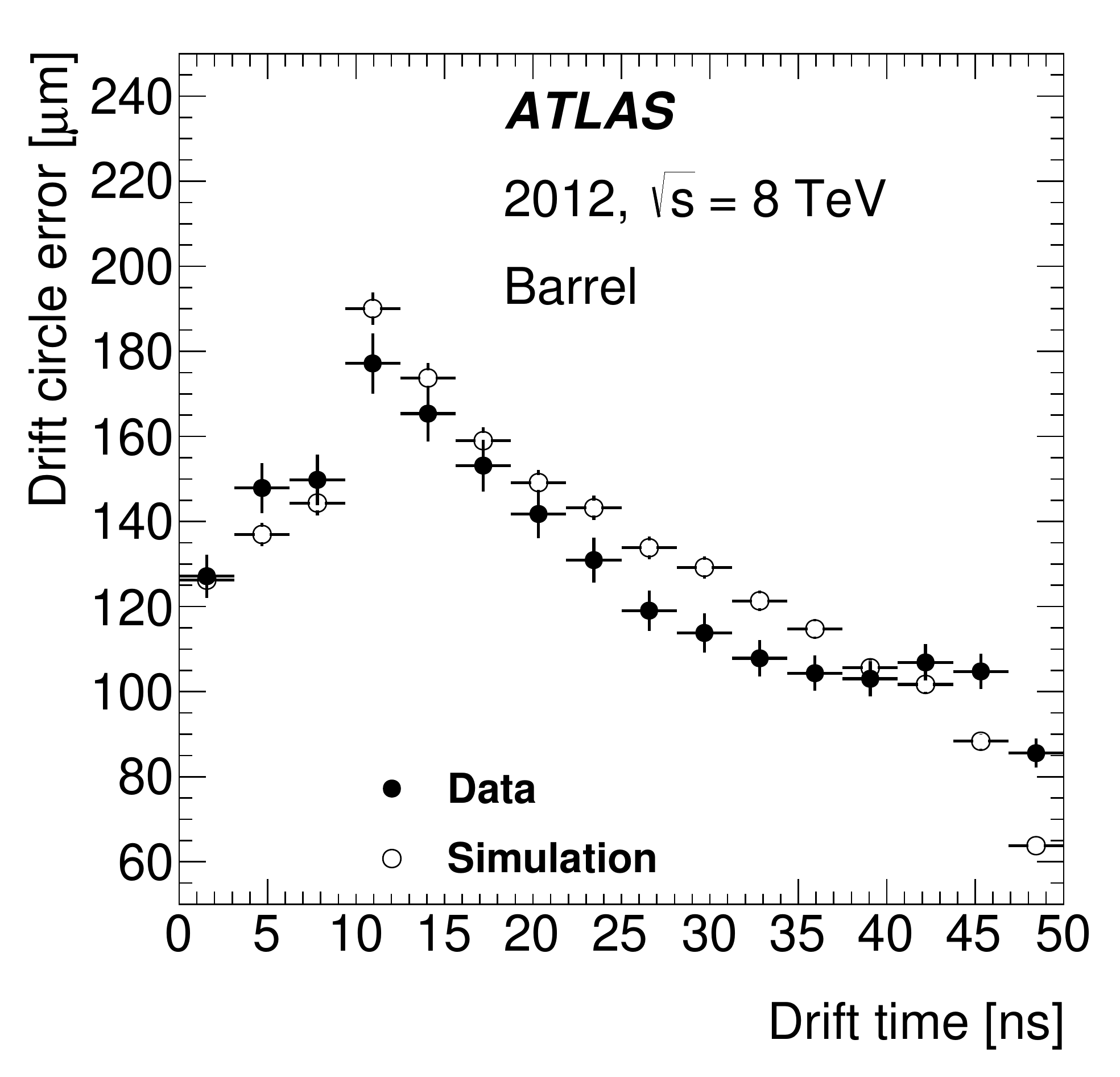}
  \label{fig:errors_vs_dt_2012_barrel}
  \includegraphics[width=0.450\columnwidth]{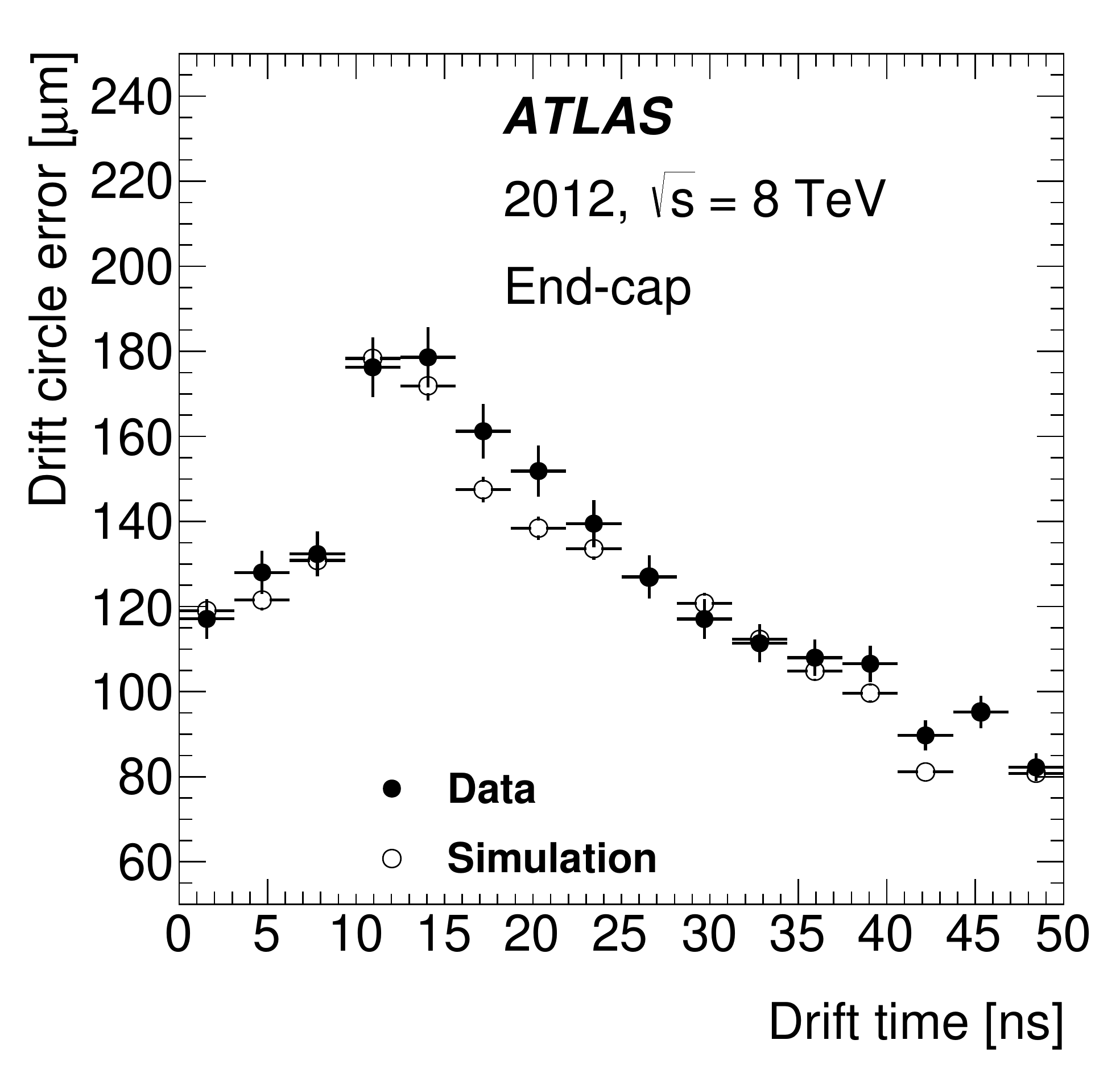}
  \label{fig:errors_vs_dt_2012_endcap} 
\caption{
Drift-circle error as a function of the drift time for the TRT barrel (left) and end-caps (right) operating with the Xe-based gas mixture for data (solid circles) and simulation (open circles).}  
\label{fig:errorf}
 \end{figure}

Figure~\ref{fig:errorf} shows typical (during Run-1 conditions with $5\le\langle\mu\rangle\le30$) drift-circle errors as a 
function of the drift time for the barrel and end-cap regions of the TRT. 
A strong dependence of the drift-circle error on the drift time is seen, which can be divided in two regions. 
For $t<10$ ns, which corresponds to the region very close to the wire (less than $\approx$300 $\mu$m), the TRT measurements  
are assigned the anode wire coordinate during the track-fitting procedure.
This procedure reduces the time jitter effect related to different arrival times of ionisation clusters at the anode wire 
as well as the effect of the left-right ambiguity (the ambiguity of the hit position being on the left or right side of the TRT straw centre) for the track-reconstruction procedure.
As a result the drift-circle error decreases for this low drift-time interval. For tracks crossing the straw at distances larger than 300 $\mu$m from the wire 
(corresponding to drift times larger than $\approx$10 ns), the drift circles are accepted as track coordinates. This leads to an increase of the drift-circle error that 
rapidly drops with increasing drift radius, due to a reduction in the fluctuations of the first ionisation cluster arrival times (the clusters reach the wires at 
almost the same times, known as a ``focusing'' effect in wire chambers).
The discrepancy between simulation and data is predominantly caused by differences in the tails of the residual distributions in data and simulation. 
Another effect contributing to the difference between simulation and data results from the approximations and models used to simulate the electrical behaviour of a 
straw tube as a transmission line. This effect is more pronounced for the barrel detector straws which are twice as long as the straws in the end-caps.

\section{Alignment}
\label{section:TRT_alignment}
The precision of momentum measurements in the ATLAS ID is determined by the intrinsic resolution of the sensitive elements, 
the level of precision with which the position in space of the detector element is known (alignment), how well the magnetic field is known, and 
the amount of material in the detector.
Misalignments in any part of ATLAS ID would degrade the momentum resolution of the reconstructed track, 
and correlated geometrical distortions can lead to systematic biases in the reconstructed track parameters.
The purpose of the alignment procedure is the precise determination of the position of the sensitive elements
as well as their geometrical distortions relative to the ideal or designed detector geometry.
A detailed description of the ATLAS ID alignment procedure and the resulting alignment accuracy are 
documented in Refs.~\cite{ATLAS-CONF-2011-012} and~\cite{ATLAS-CONF-2014-047}. Here, the basic concepts and main results relevant
for the discussion of the TRT performance are presented.

The alignment procedure calculates the TRT element positions at the level of 1\,$\mu \mathrm{m}$ precision, which is a much more precise than the typical
TRT position resolution of 110\,$\mu \mathrm{m}$. This makes the calibration procedure, which is discussed in previous section, insensitive to any small misalignments.
The determination of the position of the active elements of the TRT is performed within the ATLAS alignment framework using a 
track-based technique which minimises the position residuals. 
The alignment of each ID detector is performed at several levels of granularity. Each structure has six basic degrees of freedom, corresponding to three translations along, and three rotations about, the $x$-, $y$-, and $z$-axes, whose origin and orientation depends on the alignment level. 
First, large structures such as the full barrel or the end-caps are aligned as individual units to correct for collective movements of the active elements (Level-1 alignment). Then,
single sensitive devices are aligned (Level-2 and Levl-3 alignments). 
At Level 1, the alignment reference frame coincides with the ATLAS global coordinate frame (described in Footnote~\ref{note1}) and 
the TRT detector has the same granularity as the pixel and SCT detectors: one barrel component and two end-cap components. 

At Level 2, the alignment reference frame for the barrel modules has the $y$-axis parallel to the straw direction, the $z$-axis perpendicular to the module plane and the $x$-axis perpendicular to the $y$- and $z$-axes, forming a right-handed Cartesian system. For the end-cap wheels it originates from the centre of each wheel with the $z$-axis pointing along the beam direction, the $x$-axis being horizontal and the $y$-axis pointing upwards. At this level, the TRT barrel modules are aligned using  $x$ and $y$ translations and rotations with respect to the $x$-, $y$-, and $z$-axes. The end-cap wheels are aligned 
using only three degrees of freedom: $x$ and $y$ translations and a rotation with respect to the $z$-axis.
At this level, 620 degrees of freedom are used for the 96 barrel modules and 80 (4-plane) end-cap wheels.

Level 3 is the level where the position of each individual detector element is corrected relative to its neighbours. 
Single wires are aligned using a translation in the $\phi$ direction and a rotation around the radial direction (for the barrel) or 
an axis parallel to the beam direction and passing through the midpoint of the straw (for the end-caps). 
At this alignment level a total of 701,696 degrees of freedom are used.

As the ATLAS alignment procedure relies on the minimisation of the position residuals, the straw track position measurement accuracy
is calculated using the distribution of the unbiased residuals. The unbiased residuals are the position residuals built for tracks reconstructed
after removing all the hits from the detector element which is being aligned.
In the absence of misalignment, these position residual distributions are centred at zero, 
with their width representing the uncertainty in the drift radius measurement.
In order to converge on the best alignment for such a complex system, many iterations of the alignment procedure are performed.

\begin{center}
\begin{figure}[!ht]
\center
\includegraphics[width=0.45\textwidth]{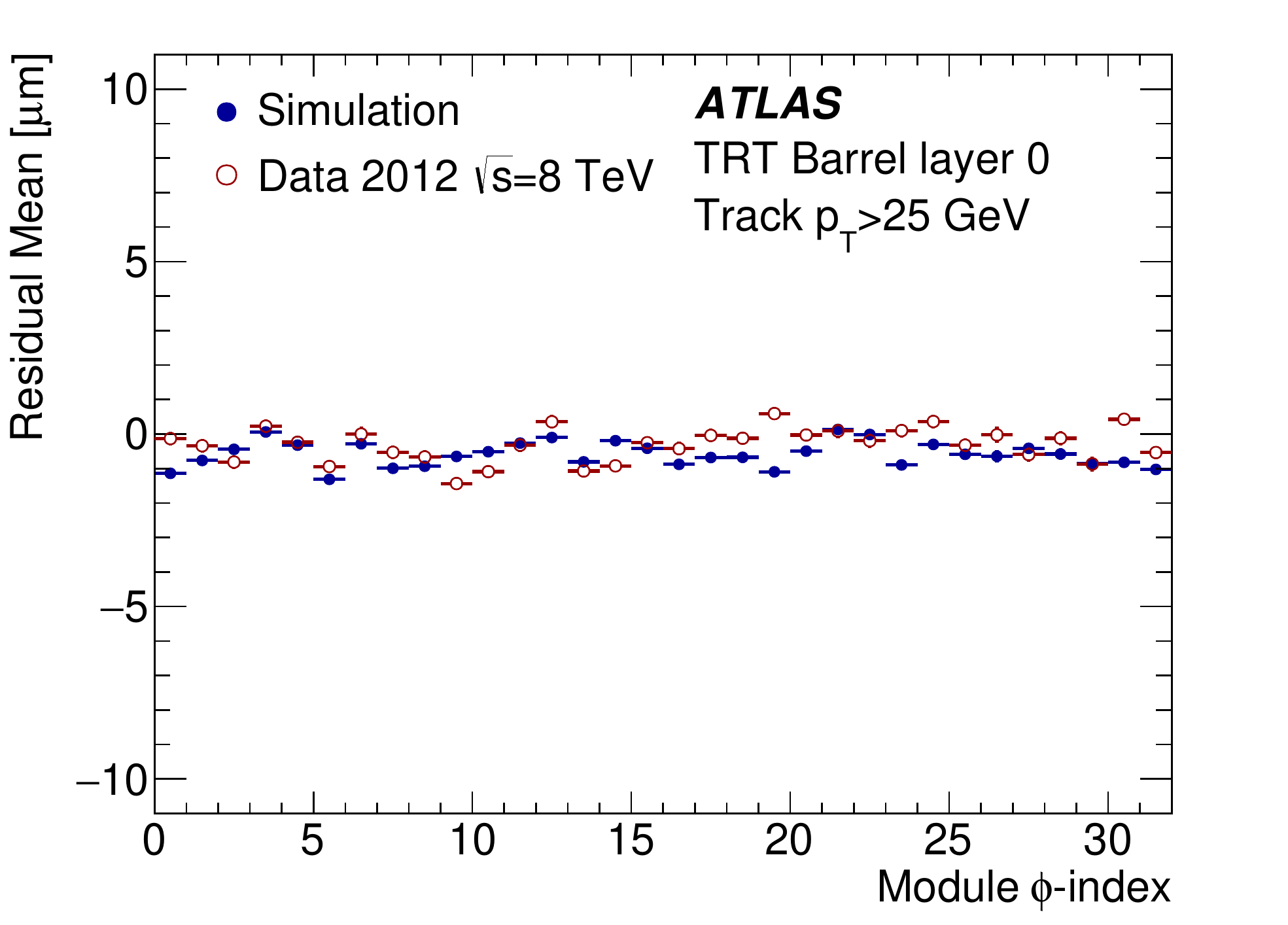}
\includegraphics[width=0.45\textwidth]{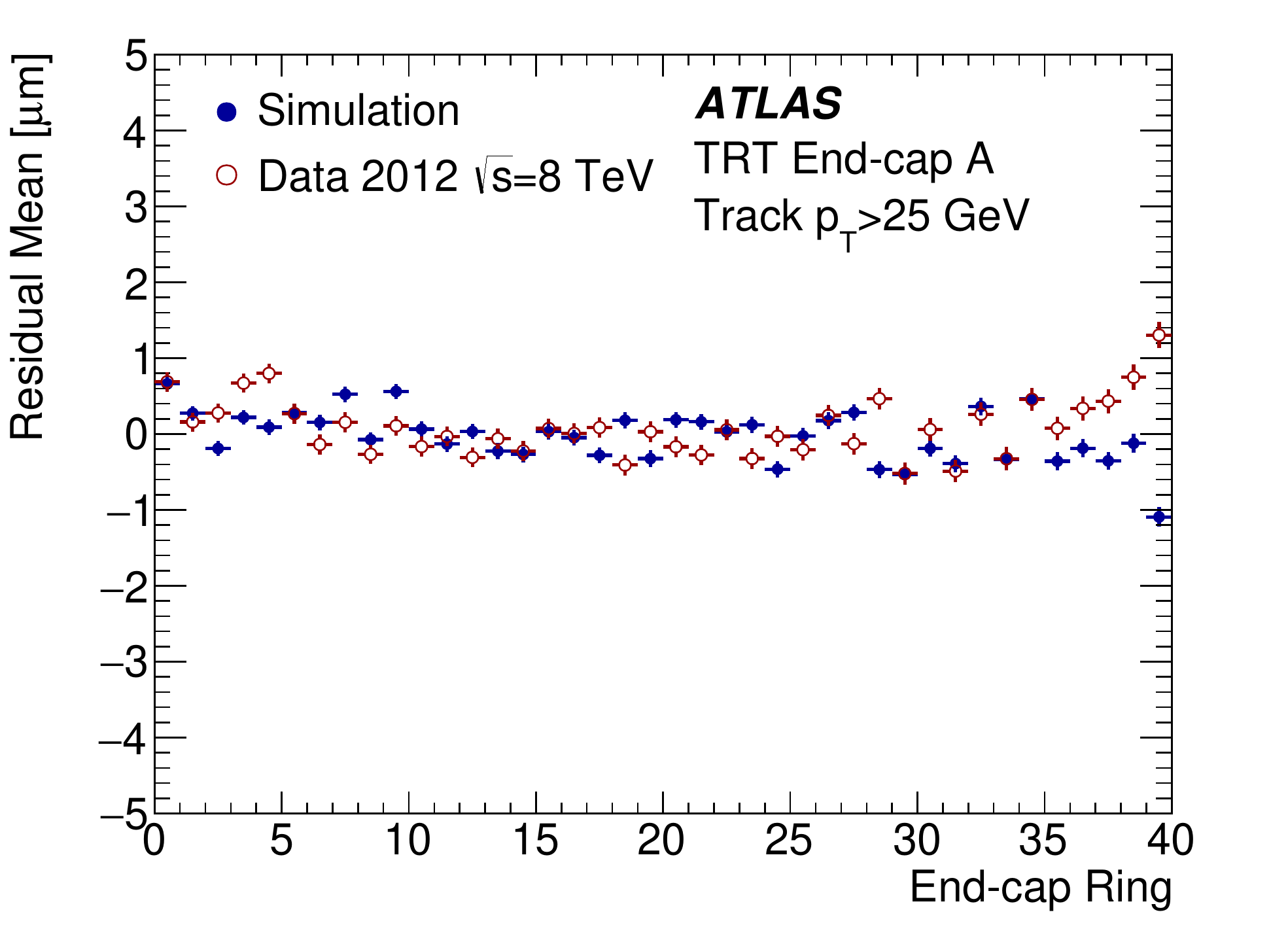}
\caption{Mean of the unbiased residual distributions as a function of the $\phi$ sector for TRT barrel layer 0 (left) and TRT end-cap A (right), operating with Xe-based gas mixture. The residuals
are from a $Z\rightarrow \mu^{+}\mu^{-}$ data sample reconstructed with the 2012 alignment constants (open circles) and are compared to the 
Monte Carlo simulation (full circles). }
\label{fig::ali2012b}
\end{figure}\end{center}

The alignment performance is checked using muon tracks from $Z$ boson decays in both data and simulation. The events are selected by requiring two muon tracks with 
$p_{\mathrm{T}}> 25~\GeV$ and  $\left|\eta\right| < 2.5$. 
In addition, the muon tracks are required to have 
at least one pixel hit and five SCT hits, and at most two pixel or SCT missing hits where they are to be expected. 
Figure \ref{fig::ali2012b} shows examples of the mean position residual distributions for the TRT barrel layer 0
and end-cap A. In the simulation a perfect alignment of the detectors is assumed. The small structures seen in the simulation position residuals are due to systematic 
mismodelling of the single detector elements' response and variation in the data-driven conditions that can affect the track-to-hit residuals.
Similar levels of agreement are observed for the other parts of the TRT.

After the alignment is performed, a typical precision on the mean of $\sim 1\,\mu \mathrm{m}$ is achieved.

\section{Performance at low occupancy}
\label{section:low_occupancy}

There are two basic straw performance parameters that define the TRT tracking properties: the straw efficiency and the straw track position measurement accuracy. 
The straw efficiency is defined as the probability for the straw to produce a signal above the LL threshold for a particle traversing
the straw gas volume. This efficiency depends on the track-to-wire distance, and is on average about 96\% for a typical straw channel.
The straw track position measurement accuracy was defined in Section~\ref{section:calib} and is typically of the order of $\sim110\, \mu \mathrm{m}$.
The individual straw performance is defined by the physics processes occurring inside the straw 
and by the signal processing in the TRT front-end electronics (described in Section~\ref{section:trt}). 

As the TRT straw drift-time parameters are measured for reconstructed 
tracks, their values depend on the track reconstruction quality, which itself is a function of the straw occupancy.
To disentangle the effect of the occupancy, baseline TRT performance studies are first performed at relatively low straw occupancy 
and compared with Monte Carlo simulations.
Low-occupancy events in data correspond to events with an average straw occupancy lower than 20\%.
 The results of the studies at low occupancy are presented in the following two subsections for Xe-based and Ar-based gas mixtures.

  \subsection{Basic tracking properties with a xenon-based gas mixture}
\label{section:xenon}

The performance of the TRT filled with a Xe-based gas mixture is studied using $pp$ collisions data taken in 2012 with a 50 ns LHC bunch spacing.
The average number of interactions per bunch crossing was in the range $5\le\langle\mu\rangle\le30$, corresponding to straw occupancies of less than 20\%.
To minimise the effects of the interaction of particles with the detector material on tracking performance, and to simplify the particle identification, the studies were carried 
out with muons from $Z$ boson decays. Events were selected in data and simulation by requiring two opposite-charge muon candidates with $\pt>30~\GeV$ and 
an angular separation between them of $\Delta R = \sqrt{(\Delta \eta)^2 + (\Delta \phi)^2} > 0.3$.
The reconstructed dimuon mass ($m_{\mu^+\mu^-}$) was required to be within $75<m_{\mu^+\mu^-}<105~\GeV$. In addition, muon tracks were 
required to fall within the TRT pseudorapidity acceptance of $|\eta|<2.0$ and to have more than 15 TRT hits.

\begin{figure}[!h]
   \begin{center}
   \includegraphics[width=0.450\textwidth]{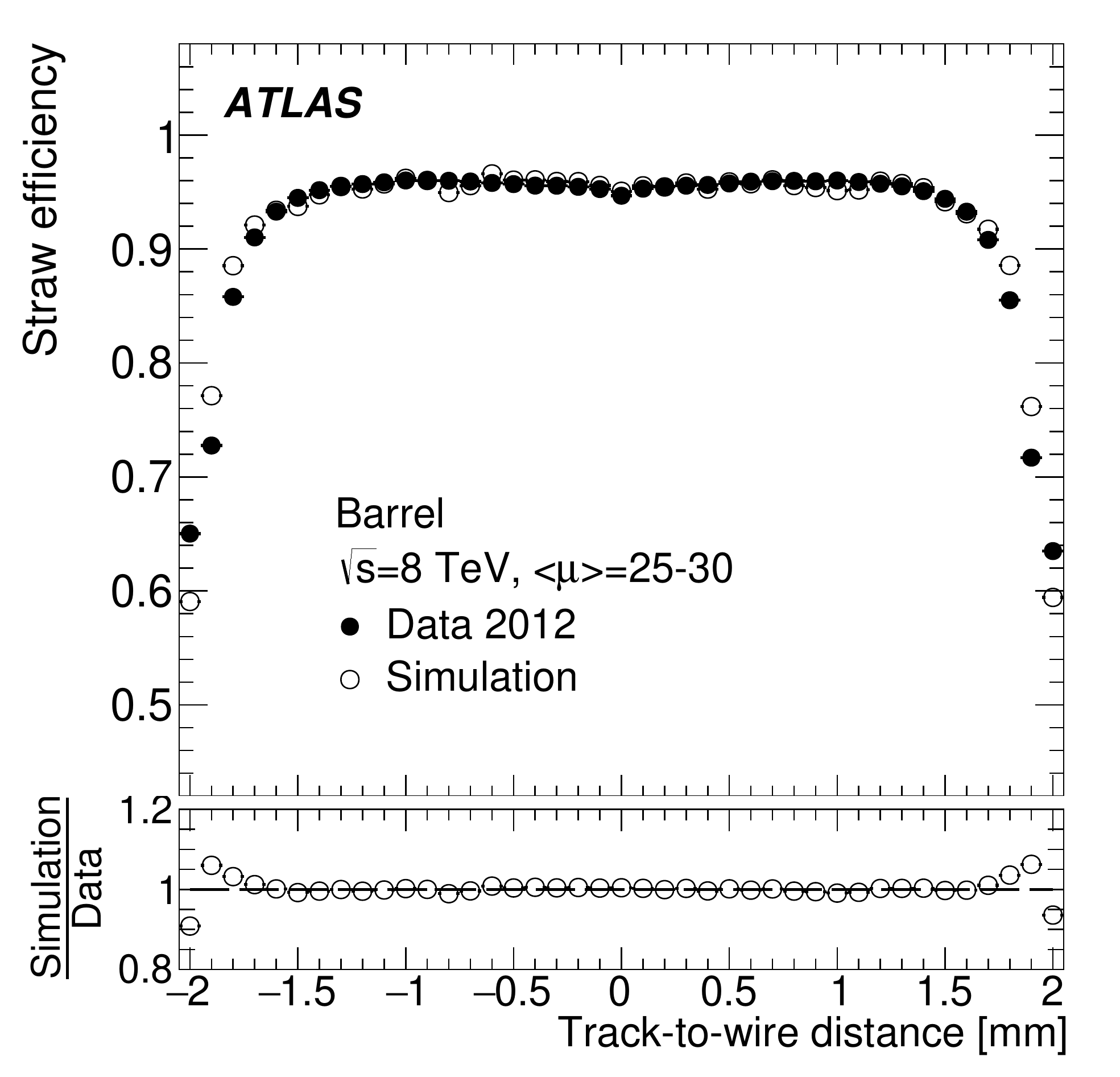} 
   \label{fig:straw_eff_t2wd_barrel_25_30_2012}
 	 \includegraphics[width=0.450\textwidth]{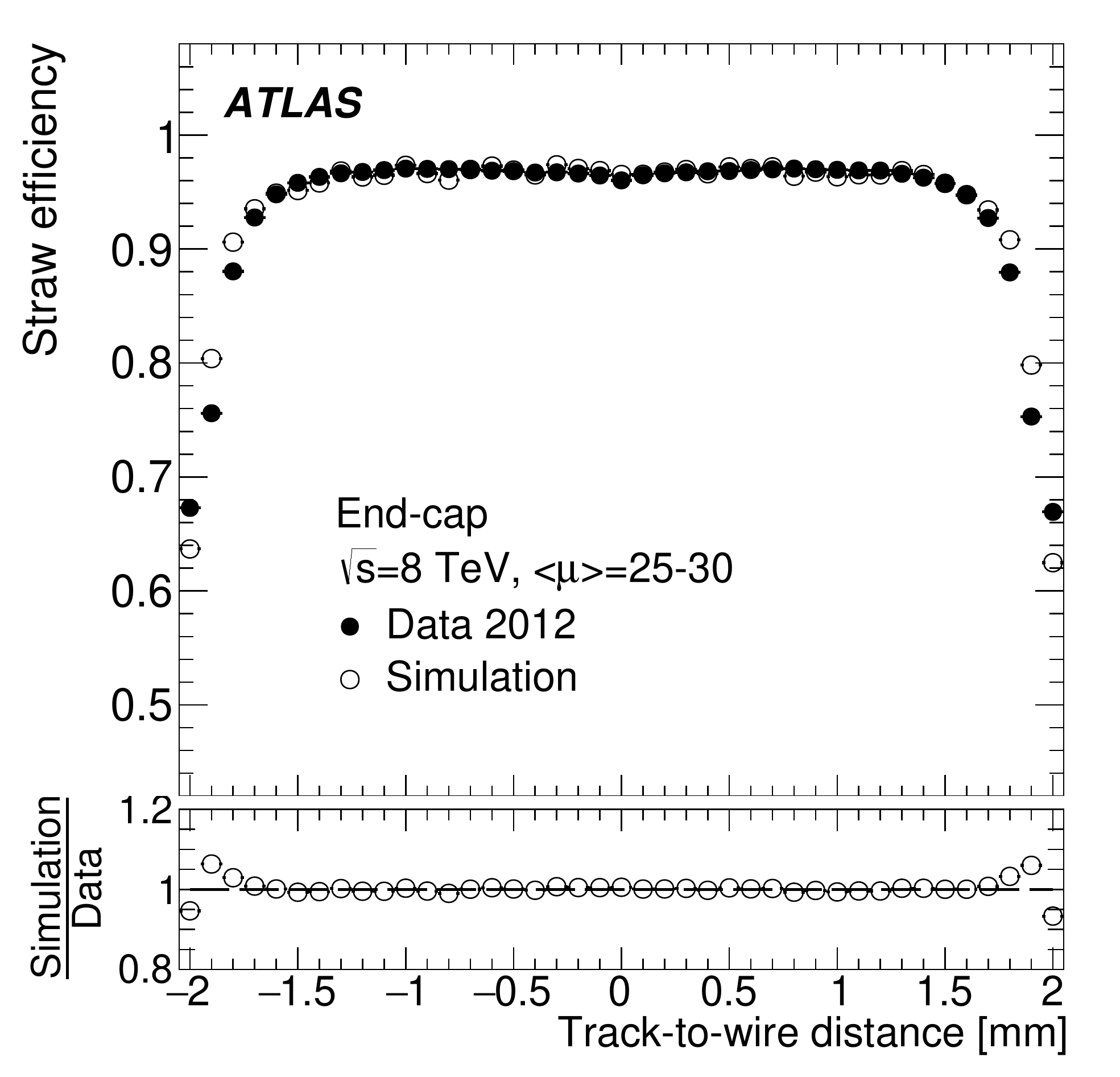} 
 	  \label{straw_eff_t2wd_ec_mu_25_30_2012}
 \caption{
  Straw efficiency in the TRT barrel (left) and end-caps (right) as a function of the track-to-wire-distance for muons with $\pt>30~\GeV$ from $Z~\rightarrow~\mu^+\mu^-$ events.
  Data (solid circles) and simulation (open circles) are shown for $25 \le \langle\mu\rangle \le 30$ during the 2012 running period operating with the Xe-based gas mixture at $\sqrt{s} = 8~\TeV$.}
  \label{fig:straw_eff_t2wd_mu_25_30_2012}
    \end{center}
\end{figure}

A comparison of the straw efficiency from  $Z \rightarrow \mu^+\mu^-$ data and simulation is shown in
Figure~\ref{fig:straw_eff_t2wd_mu_25_30_2012} as a function of the track-to-wire distance for the barrel and end-caps. 
One can see a plateau at about 96\% in straw efficiency for absolute values of track-to-wire distance below 1.5\,mm, and a significant decrease in the region where the track passes close to the edge of the straw, due to the shorter track length in the straw.
As mentioned earlier, the TRT LL threshold is set according to the TRT noise occupancy requirements and its absolute value in terms of energy scale is not a priori exactly known. 
Consequently, the LL threshold  setting in the simulation was tuned to obtain agreement with the data for the plateau region of the straw efficiency distributions, separately for
the barrel and end-caps. 
Agreement is best when the effective LL threshold is set to about $280~\eV$ for the TRT barrel and to about $250~\eV$ for the end-caps.

\begin{figure}[!h]
  \centering
  \includegraphics[width=0.450\columnwidth]{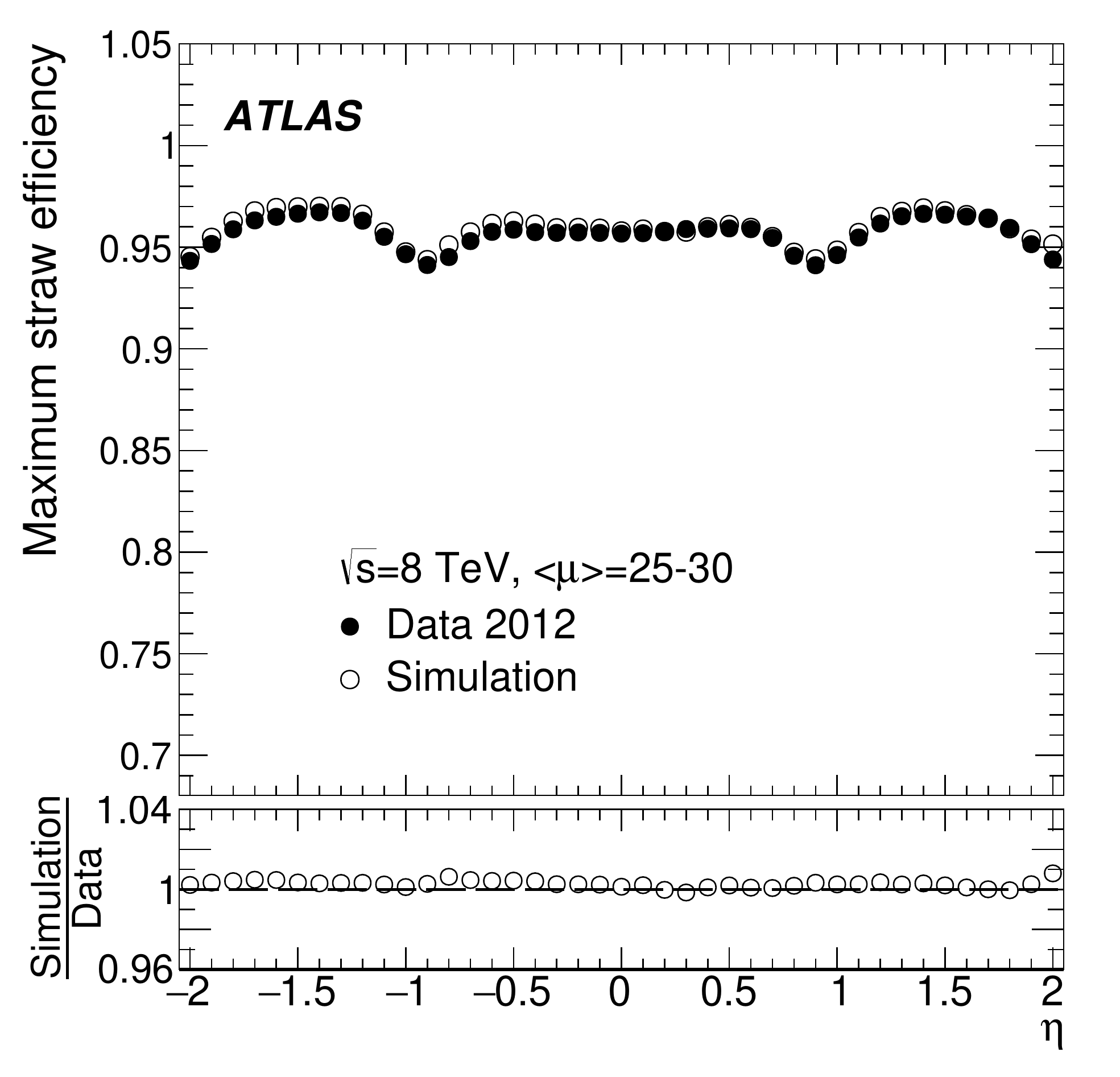} 
  \caption{
  Maximum straw efficiency as a function of $\eta$ for muons with $\pt>30~\GeV$ from $Z~\rightarrow~\mu^+\mu^-$ events.
  Data (solid circles) and simulation (open circles) are shown for $25 \le \langle\mu\rangle \le 30$ during the 2012 running period operating with the Xe-based gas mixture at $\sqrt{s} = 8$\,TeV.}
  \label{fig:straw_eff_eta_mu_25_30_2012}
 \end{figure}

Figure~\ref{fig:straw_eff_eta_mu_25_30_2012} shows the maximum straw efficiency as a function of the track $\eta$. 
In this figure, the maximum straw efficiency is extracted from a constant fit to the plateau of the straw efficiency vs track-to-wire distance for each bin of $\eta$. 
The maximum straw efficiency is found to be 96\% on average. The few percent variations around this value are related to the variation of the particle incident 
angle relative to the straw and to the change of the signal amplitude along the wire. 
The straw efficiency is smaller for particle crossing the straws close to the barrel edge  (in the barrel to end-cap transition region) because the number of straws crossed by particles in this region is smaller and because this is where the services for the SCT and TRT stands, along with the TRT barrel electronics. This leads to some reduction of the track reconstruction accuracy and hence decrease of the straws efficiency.

\begin{figure}[!h]
  \centering
  \includegraphics[width=0.450\columnwidth]{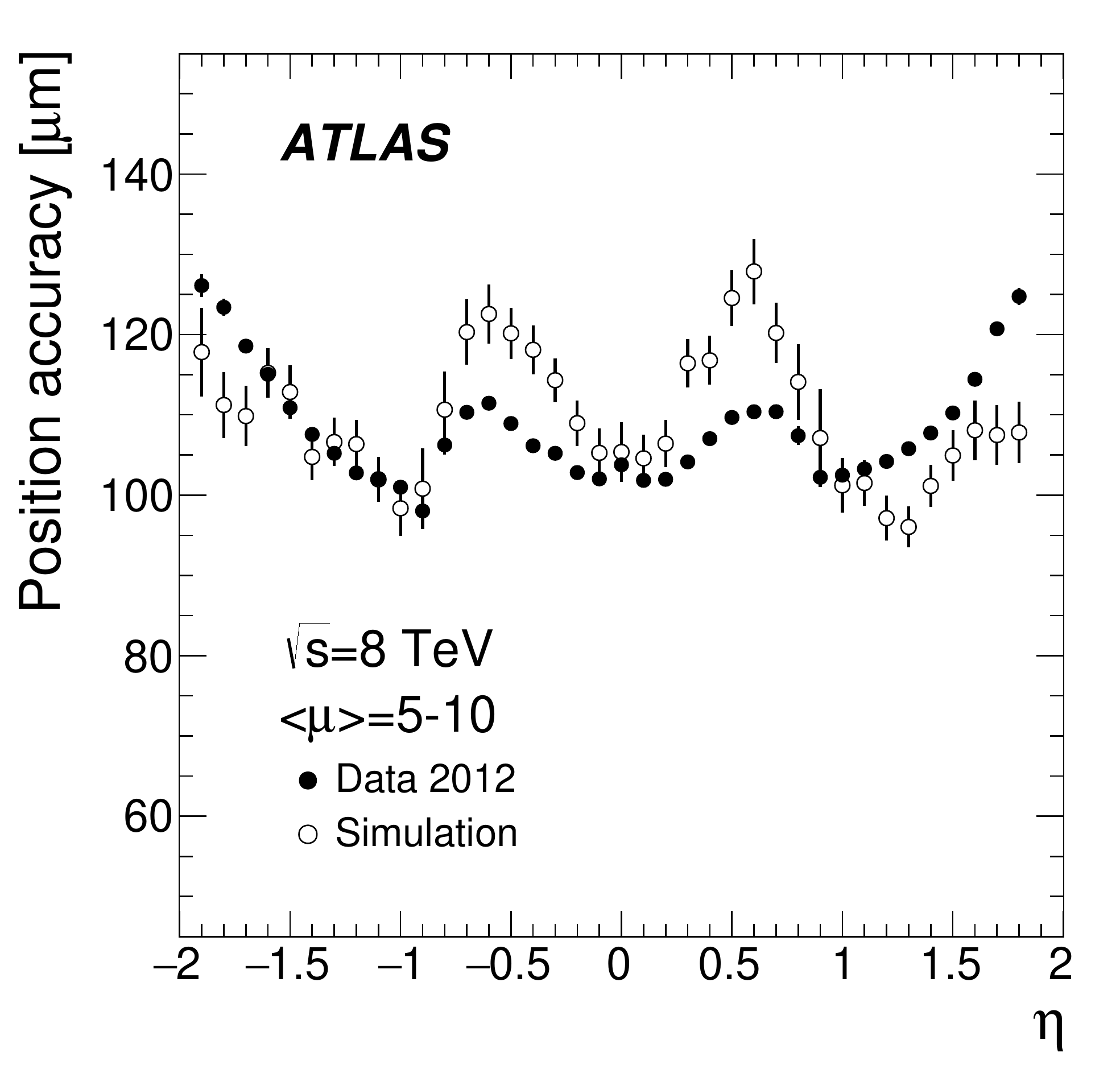}
  \caption{
  Straw track position measurement accuracy as a function of $\eta$ for muons with $\pt>30~\GeV$ from $Z~\rightarrow~\mu^+\mu^-$ events.
  Data (solid circles) and simulation (open circles) are shown for the 2012 running period operating with Xe-based gas mixture with $5\le\langle\mu\rangle\le10$.}
  \label{fig:hit_res_eta_unbiased_mu_5_10-ref} 
 \end{figure}

Figure~\ref{fig:hit_res_eta_unbiased_mu_5_10-ref}  shows the straw track position measurement accuracy as a function of  $\eta$ for data and simulation. 
Good agreement between simulation and data was found for the end-cap region of the TRT, while for the barrel the simulation predicts a smaller position accuracy than observed in the data (up to about 10\,$ \mu \mathrm{m}$). 
As described in Section \ref{section:calib} for Figure \ref{fig:errorf}, this different degree of agreement between data and simulation in the barrel and the end-cap regions results from approximations and models used to simulate the electrical behavior of a straw tube as a transmission line. For the barrel detector, straws are twice as long as the straws in the end-caps.
As already observed for the straw efficiency, there is a variation of the straw track position measurement accuracy around the central value of $\sim$110\,$ \mu \mathrm{m}$ for the same reasons as mentioned earlier. In particular, the change of the signal amplitude along the wire has a significant effect for the long straws in the barrel.

\begin{figure}[!h]
  \centering
\includegraphics[width=0.45\columnwidth]{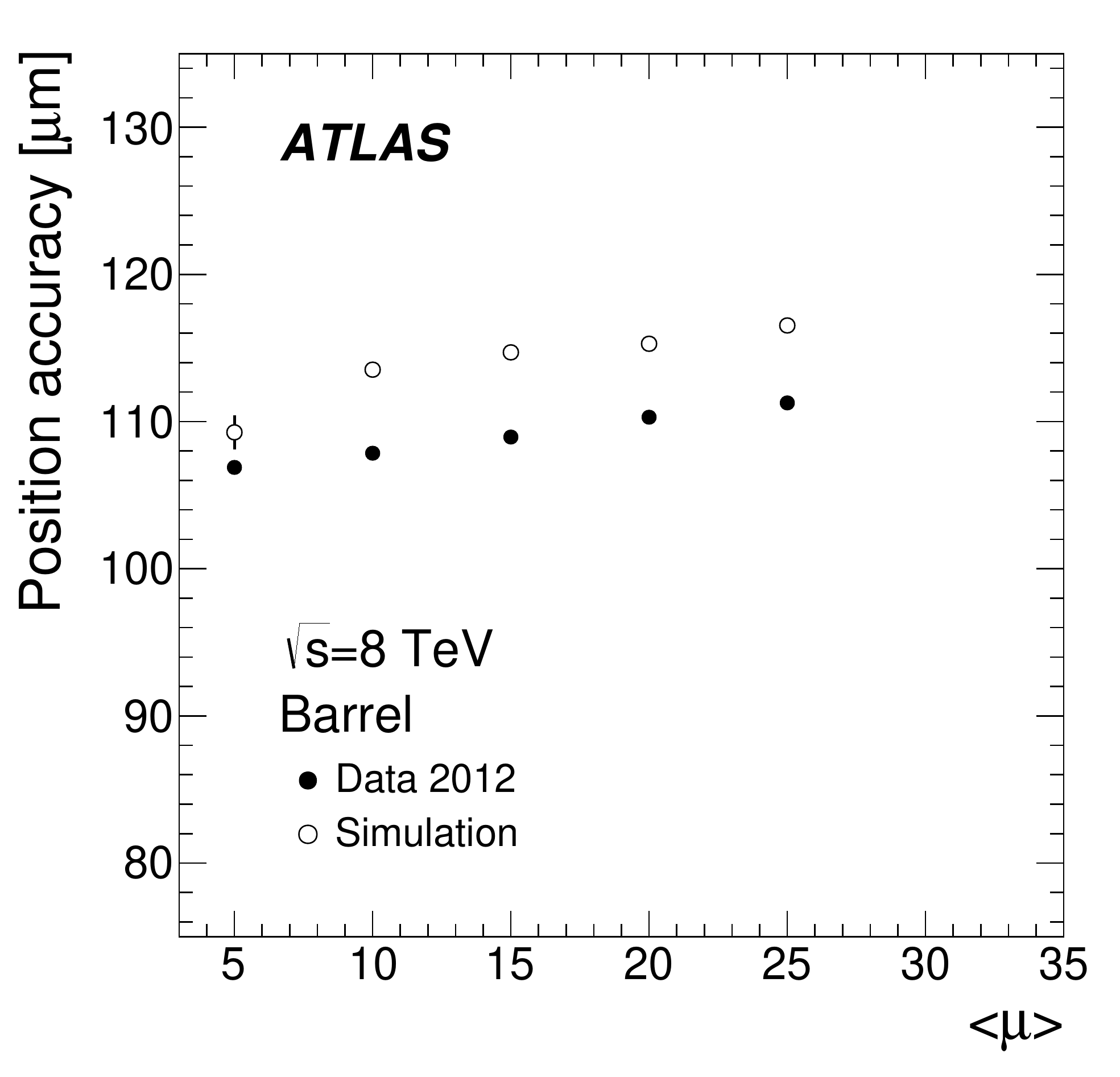}
  \label{fig:hit_res_mu_unbiased_barrel-ref}
  \includegraphics[width=0.45\columnwidth]{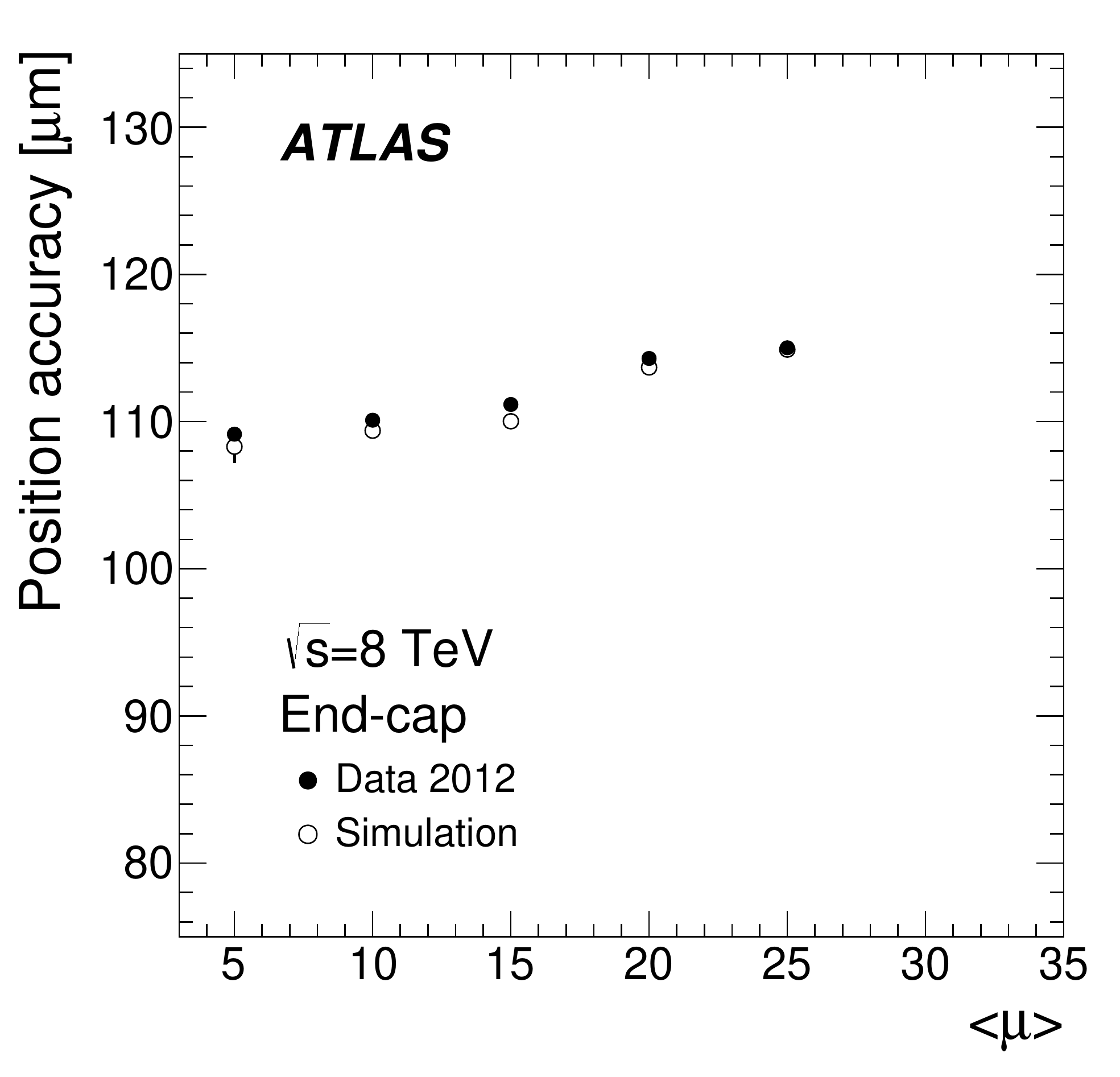}
  \label{fig:hit_res_mu_unbiased_ec-ref}
  \caption{
    Straw track position measurement accuracy for muons with $\pt>30~\GeV$ from $Z~\rightarrow~\mu^+\mu^-$ events in the TRT barrel (left) and end-caps (right) as a function of $\langle\mu\rangle$.  Data (solid circles) and simulation (open circles) are shown for the 2012 running period operating with a Xe-based gas mixture and for $5\le\langle\mu\rangle\le30$.} 
  \label{fig:hit_res_mu_unbiased-ref}
 \end{figure}

Although low-occupancy LHC fills were chosen for these studies ($5\le\langle\mu\rangle\le30$) there is still some dependence of the straw track 
position measurement accuracy on occupancy. 
Figure~\ref{fig:hit_res_mu_unbiased-ref} shows the averaged straw track position measurement accuracy as a function of $\langle\mu\rangle$ for the TRT  barrel and end-caps, comparing data with simulation.
As seen in Figure~\ref{fig:hit_res_eta_unbiased_mu_5_10-ref}, the simulation underestimates the straw track position measurement accuracy in  the barrel straws
by about $6\,\mu \mathrm{m}$ on average, but the general trend as a function of $\langle\mu\rangle$ is well reproduced.
For the end-caps the simulation is in agreement with data.
In the data, the average straw track position measurement accuracy over the entire TRT is in the range $105~ \mu \mathrm{m}$--$115~\mu \mathrm{m}$. 

  \subsection{Basic tracking properties with an argon-based gas mixture}
\label{section:argon}

An Ar-based gas mixture was considered as an alternative to the Xe-based gas mixture for the TRT operation at the detector design stage.
While it is not appropriate for electron identification due to its inability to absorb TR efficiently, it is expected to be fully adequate 
for tracking purposes. 
A special option for signals from an Ar-based gas mixture in the front-end signal-shaping electronics was foreseen in the TRT read-out design~\cite{1748-0221-3-06-P06007}. 
Because of the specific signal shape from the straw for an Ar-based mixture, the TRT read-out electronics integrates more
charge during the signal processing than for a Xe-based mixture (about 15\% instead of 5\%) 
and hence can operate at a lower effective threshold. A lower threshold should result in an improvement
of the drift-time measurement accuracy.
On the other hand, the Ar-based mixture has less density than the corresponding Xe-based one, leading to an increase in 
the average distance between ionisation clusters and hence to a degradation of the drift-time measurement accuracy. 
In order to test the performance of an Ar-based mixture, and to be able to tune the simulation,
a gas mixture of 70\% Ar, 27\% $\mathrm{CO_2}$, and 3\% $\mathrm{O_2}$ was used to fill the straws in some parts of the TRT (four modules in the first layer of the barrel and the 
end-cap wheels number 7 and 8 on the negative $\eta$ side, see Figure~\ref{fig:AtlasInnerDetector}) during the proton--lead collision data-taking period in 2013.
Different runs with various high-voltage settings were taken for the Ar-filled straws.
In order to compare data with simulation, a LL threshold scan in the simulation was performed to match the different high-voltage settings in data.

\begin{figure}[!h]
  \centering
  \includegraphics[width=0.460\columnwidth]{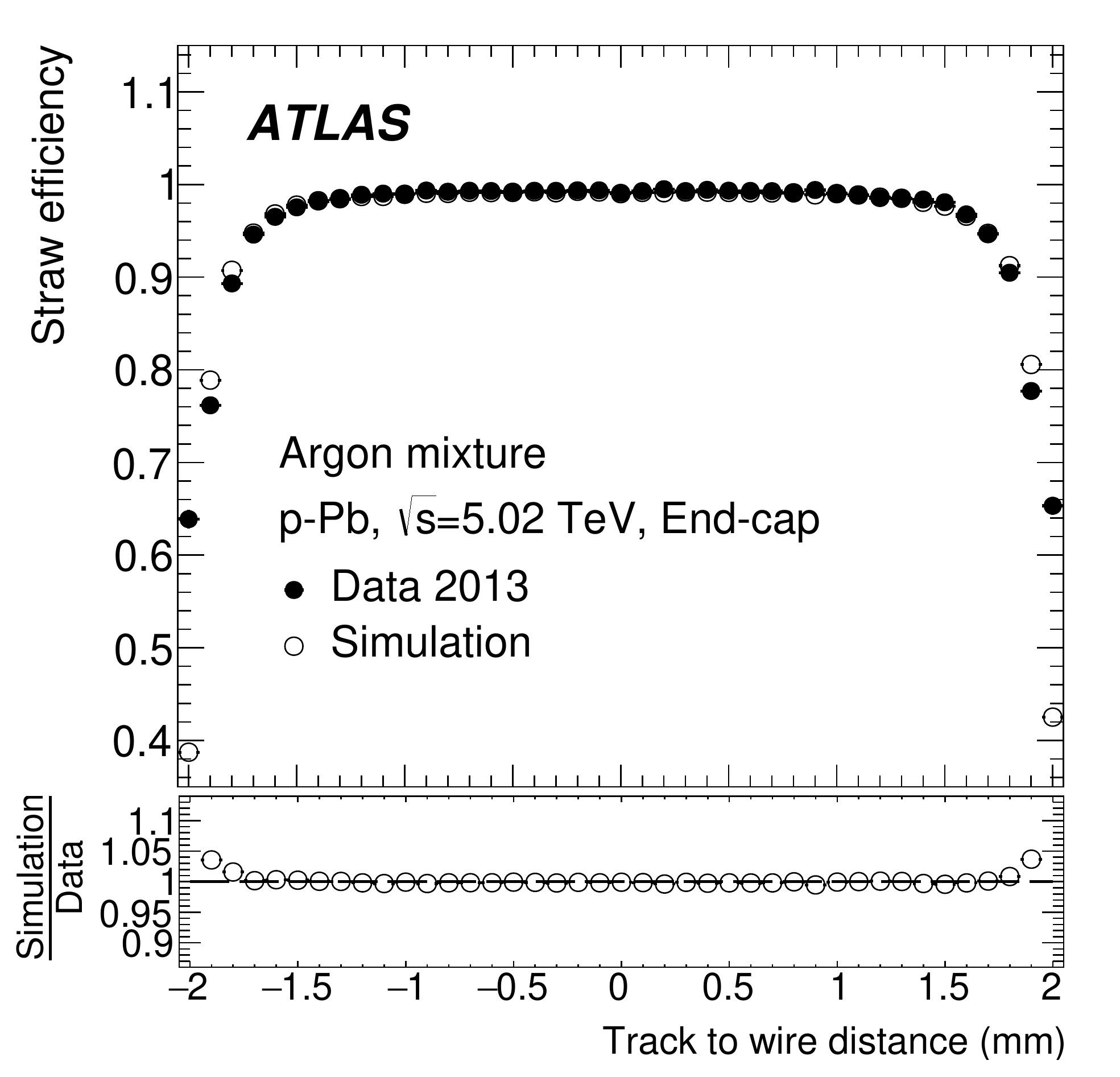}
  \caption{
  Straw efficiency in the end-cap wheels filled with an Ar-based mixture versus track-to-wire distance
 in the data (solid circles) and simulation (open circles).}
\label{fig:FigArgonSEvT2WD}
 \end{figure}

Figure~\ref{fig:FigArgonSEvT2WD} shows the straw efficiency for an Ar-based mixture in the end-cap wheels as a function of the track-to-wire 
distance. This figure shows the same behaviour as seen previously for the Xe-based gas mixture (see Section~\ref{section:xenon}), with a plateau 
close to 100\% efficiency and agreement between simulation and data. The higher plateau found here reflects the fact that 
for argon, the effective threshold is about a factor of three lower than that for xenon, thus increasing  
the sensitivity of the read-out electronics to low ionisation-energy clusters.

\begin{figure}[!h]
  \centering
  \includegraphics[width=0.460\columnwidth]{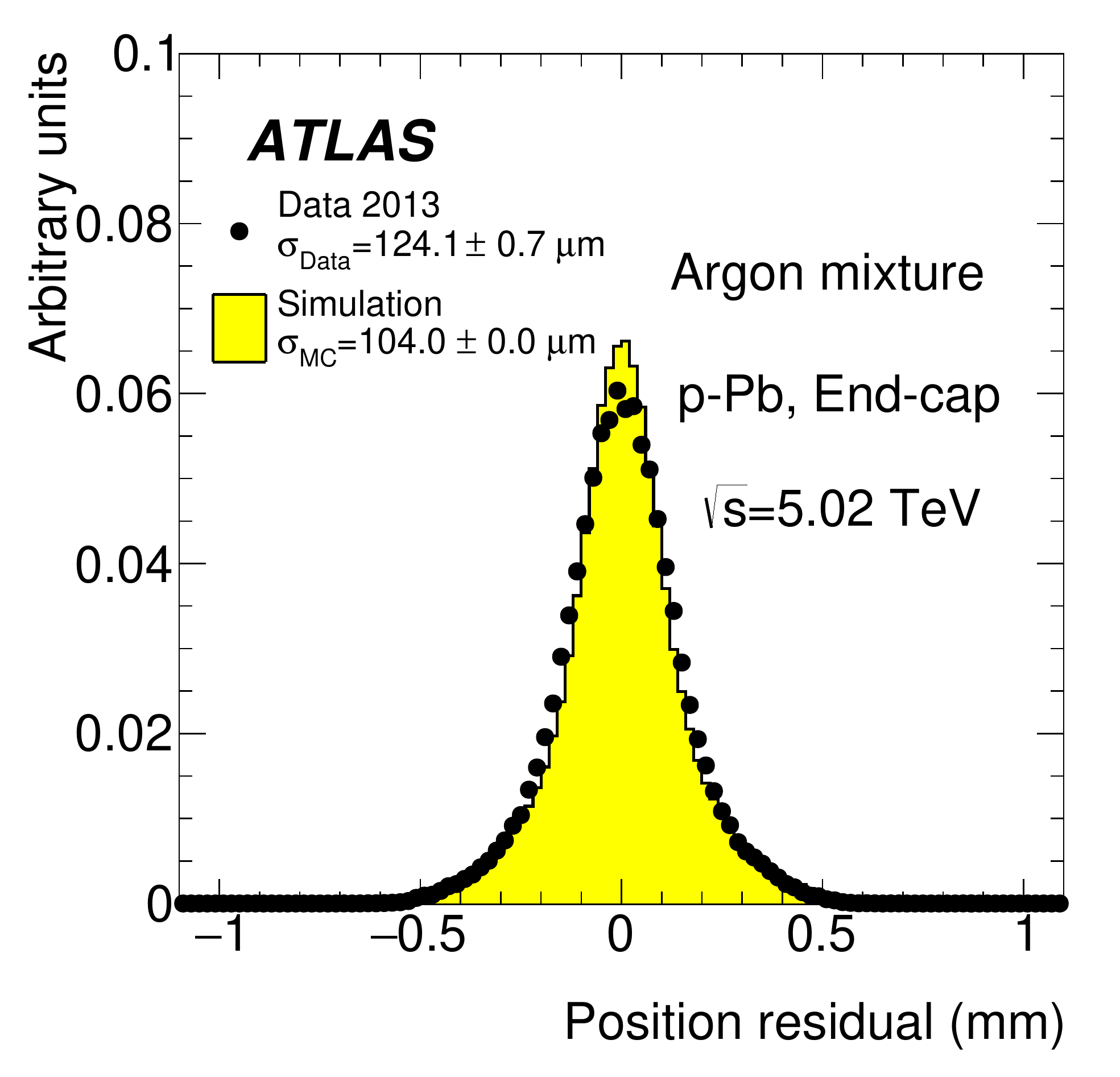}
  \caption{
Position residual distribution for the Ar-based end-cap straws for data and     
simulation using the $100~\eV$ threshold setting.}
\label{fig:FigArgonPRvT2WD}
 \end{figure}

Figure~\ref{fig:FigArgonPRvT2WD} depicts an example of the track position residual distribution for the Ar-filled straws in end-cap, for data and 
simulation, and for a LL threshold setting of about $100~\eV$. 
The simulation predicts better accuracy than that observed in data. This points to a small mismodelling of 
some parameters for the Ar-based mixture. 

\begin{figure}[!h]
  \centering
  \includegraphics[width=0.460\columnwidth]{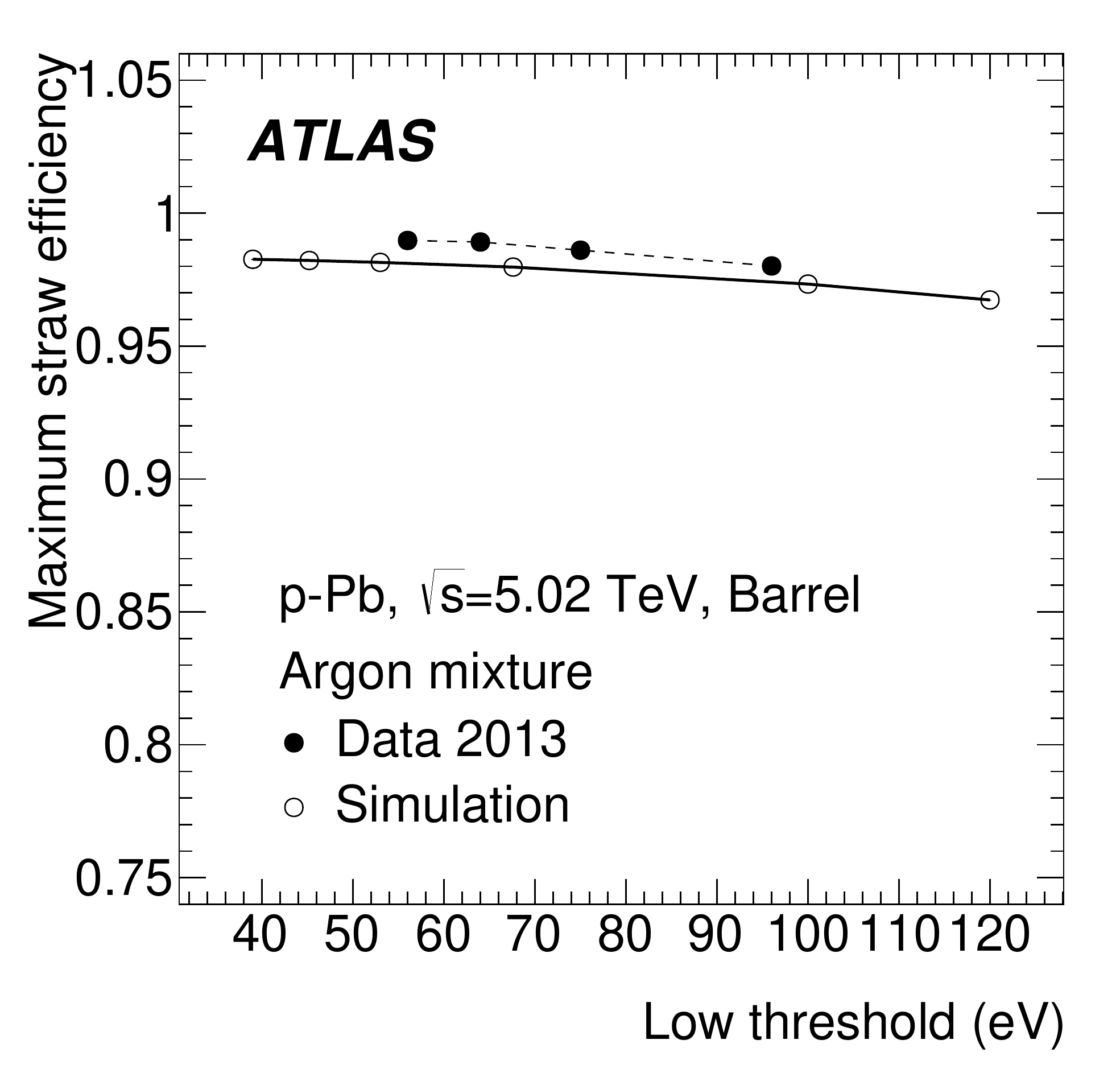}
  \label{fig:P0_StrawEff_T2WD_Barrel_LongArg}
  \includegraphics[width=0.460\columnwidth]{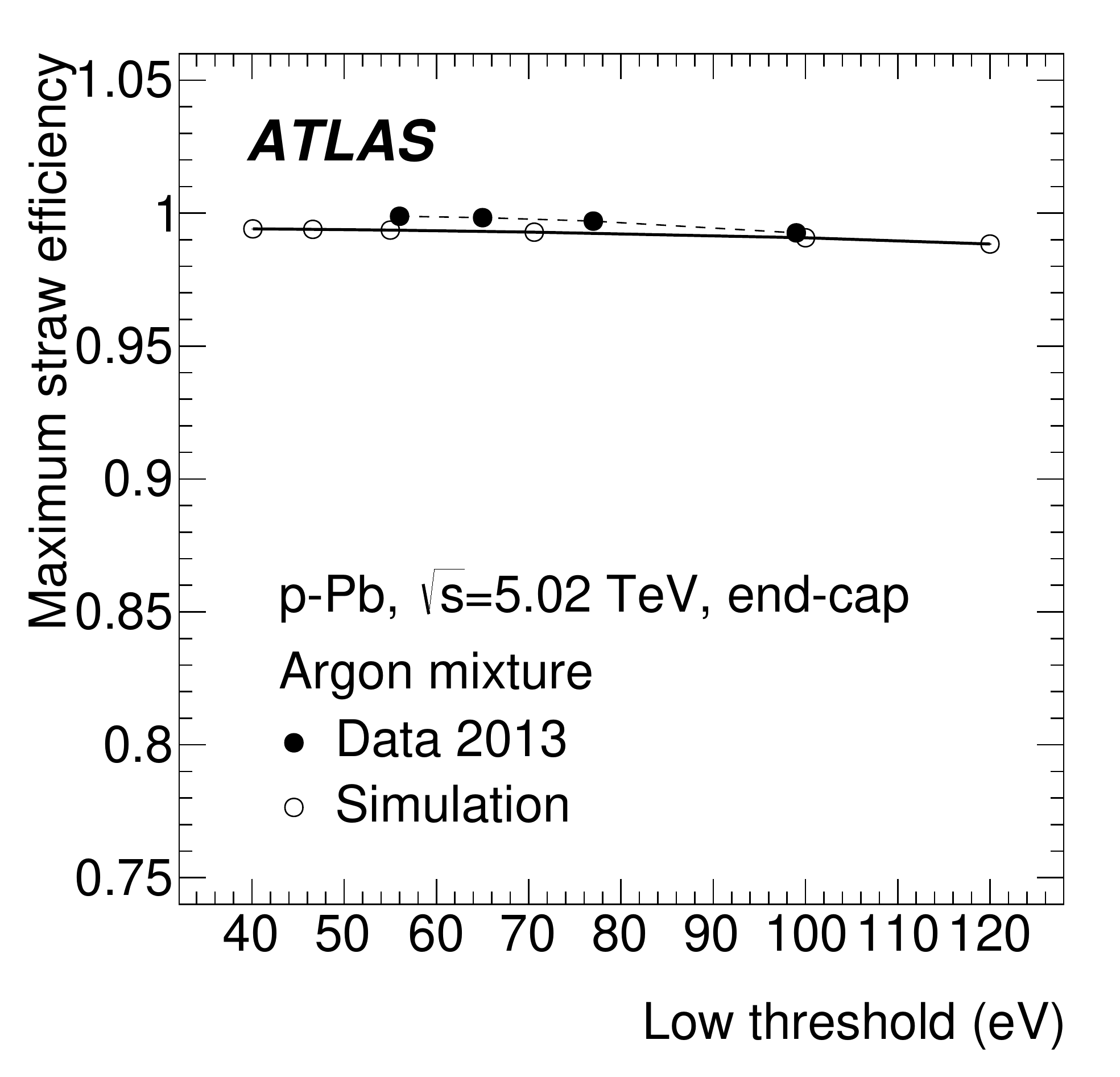}
  \label{fig:P0_StrawEff_T2WD_Barrel_EndcapCArg}
  \caption{
  Maximum straw efficiency as a function of the threshold setting for the barrel modules (left) and end-cap wheels (right) filled with an Ar-based gas mixture, for 
data (solid circles) and simulation (open circles).} 
\label{fig:FigArgonSEvLT}
 \end{figure}

\begin{figure}[!h]
  \centering
  \includegraphics[width=0.460\columnwidth]{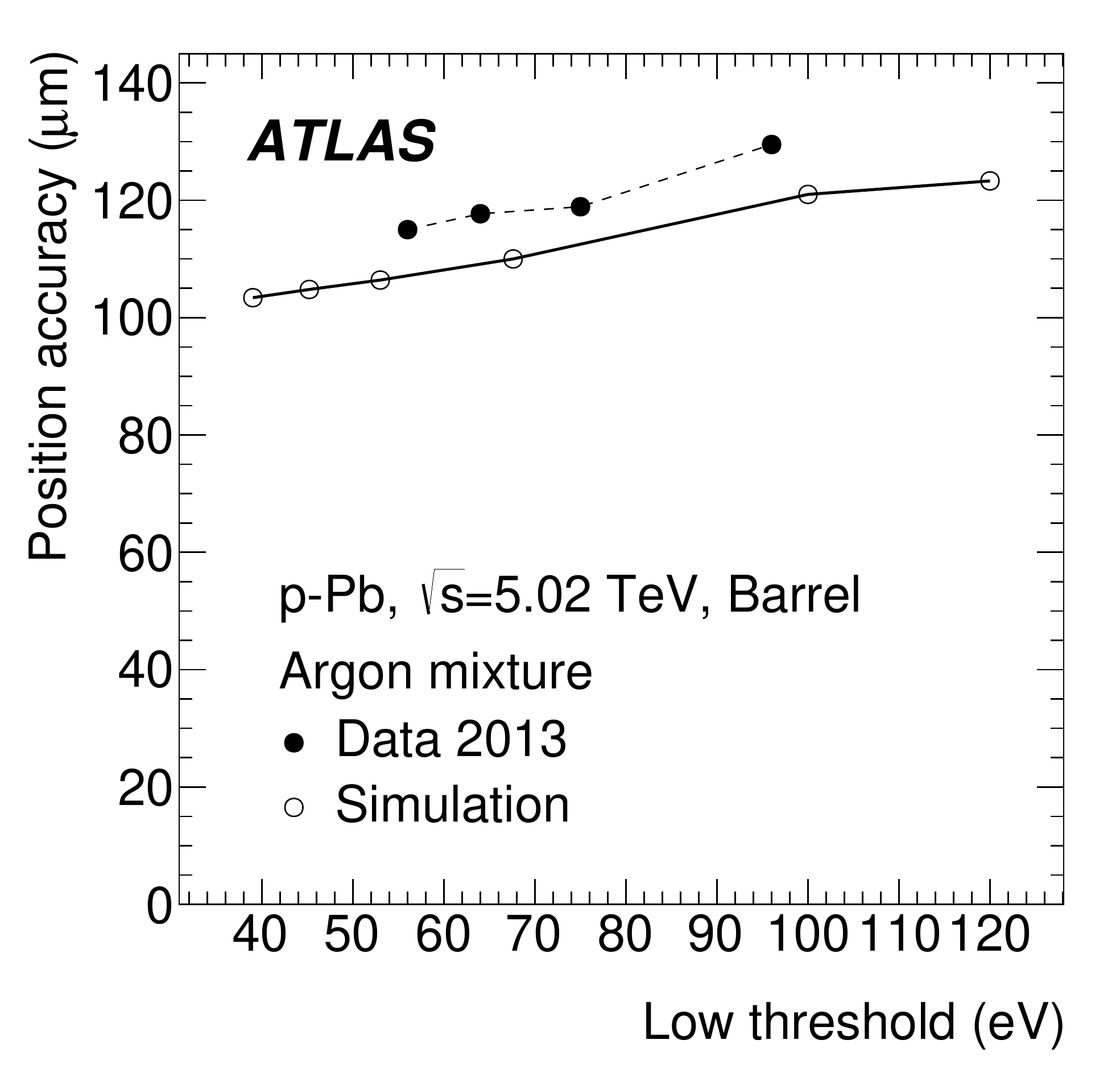}
  \label{fig:Sigma_PreciseHits_Barrel_Lay0_Long_Arg_data_MC}
  \includegraphics[width=0.460\columnwidth]{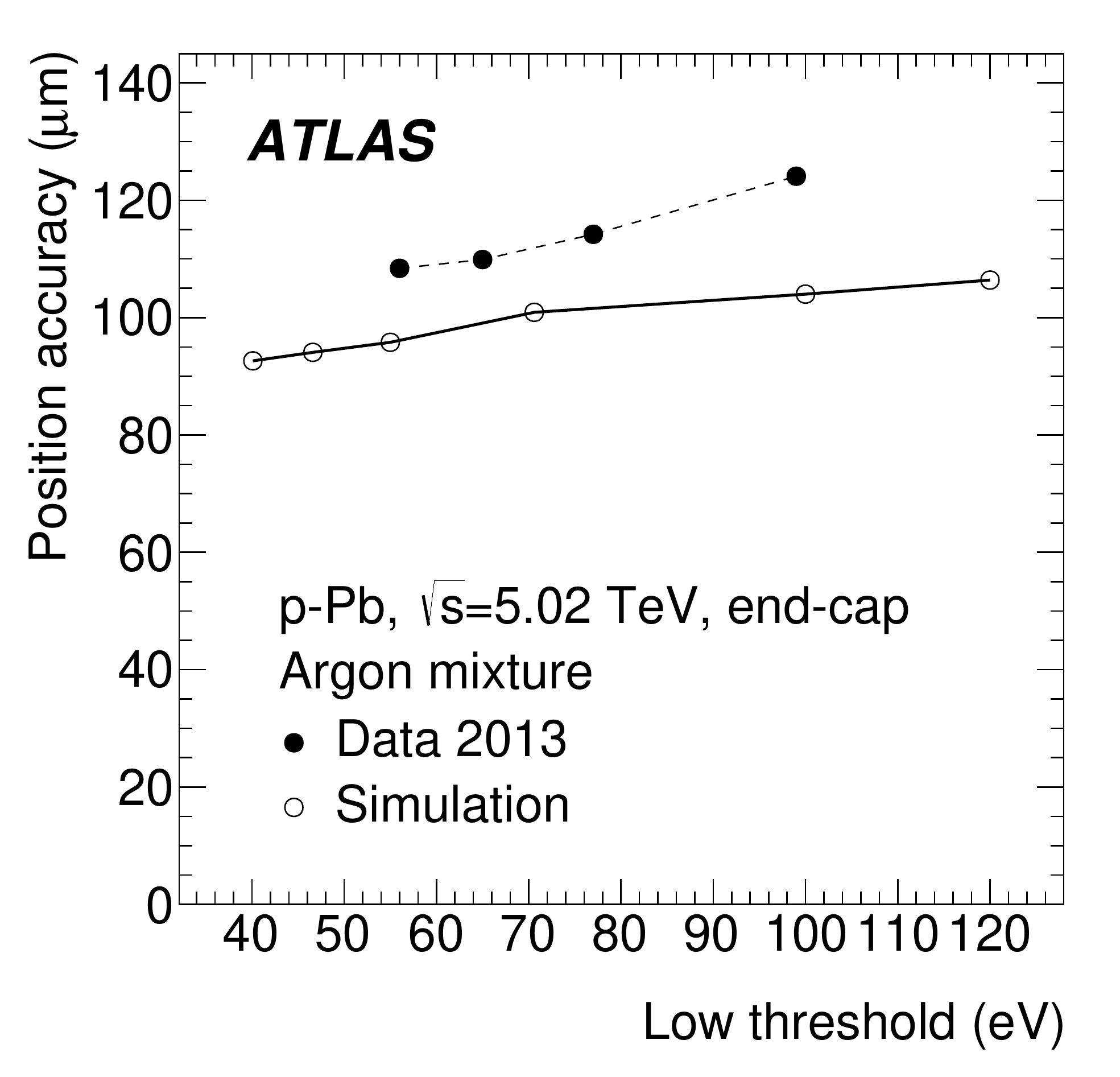}
  \label{fig:Sigma_PreciseHits_EndcapC_Arg_data_MC}
  \caption{
    Straw track position measurement accuracy as a function of the threshold setting for the barrel modules (left) and end-cap wheels (right) filled with an
  Ar-based gas mixture, for the data (solid circles) and simulation (open circles).} 
\label{fig:FigArgonPRvLT}
 \end{figure}

The maximum straw efficiency and straw track position measurement accuracy  as a function of  different LL threshold settings are shown in Figures~\ref{fig:FigArgonSEvLT} and \ref{fig:FigArgonPRvLT} for the barrel and the end-caps.
The straw efficiency shows little sensitivity to the LL threshold in the tested range
whilst the dependence is more significant for the straw track position measurement accuracy.
There is good agreement between simulation and data for the straw efficiency, 
while the straw track position accuracy is found to be about 10--15\% worse in data. 
The straw track position accuracy is below $125\,\mu \mathrm{m}$ for all thresholds and over the entire TRT.

\section{Tracking properties at high occupancy}
\label{section:occupancy}

During the Run 1 data-taking period, the standard bunch spacing was set to 50~ns whereas the nominal LHC running condition was designed with 25~ns intervals between collisions. 
With the 50~ns spacing, the occupancy in $pp$ collisions reached in the TRT was at the level of $\sim$25\%, and approached the 90\% level in heavy ion collisions.
During the Run-2 period, started in 2015, the LHC operates with the design 25~ns bunch spacing and provides higher instantaneous luminosity. In these conditions the TRT straw occupancy significantly increases not only due to the increased luminosity but also due to the contribution from additional collisions in neighbouring bunches (out-of-time pile-up), which become more probable with the reduced bunch spacing.

The TRT is designed to operate in a high occupancy environment thanks to the large number of measurements (30) provided by the TRT along reconstructed tracks. 
Even at very high occupancy there are always some precision hits which significantly contribute to the track-finding algorithm and 
to the particle momentum reconstruction accuracy. High-occupancy conditions are studied to understand the effect of large charged-particle densities 
on the TRT performance. The studies were carried out using a special high-$\langle\mu\rangle$ LHC fill during a $pp$ data-taking period in 2012, 
and using data from heavy ion collisions taken in 2011. 

\begin{figure}[!h]
  \centering
  \includegraphics[width=0.450\columnwidth]{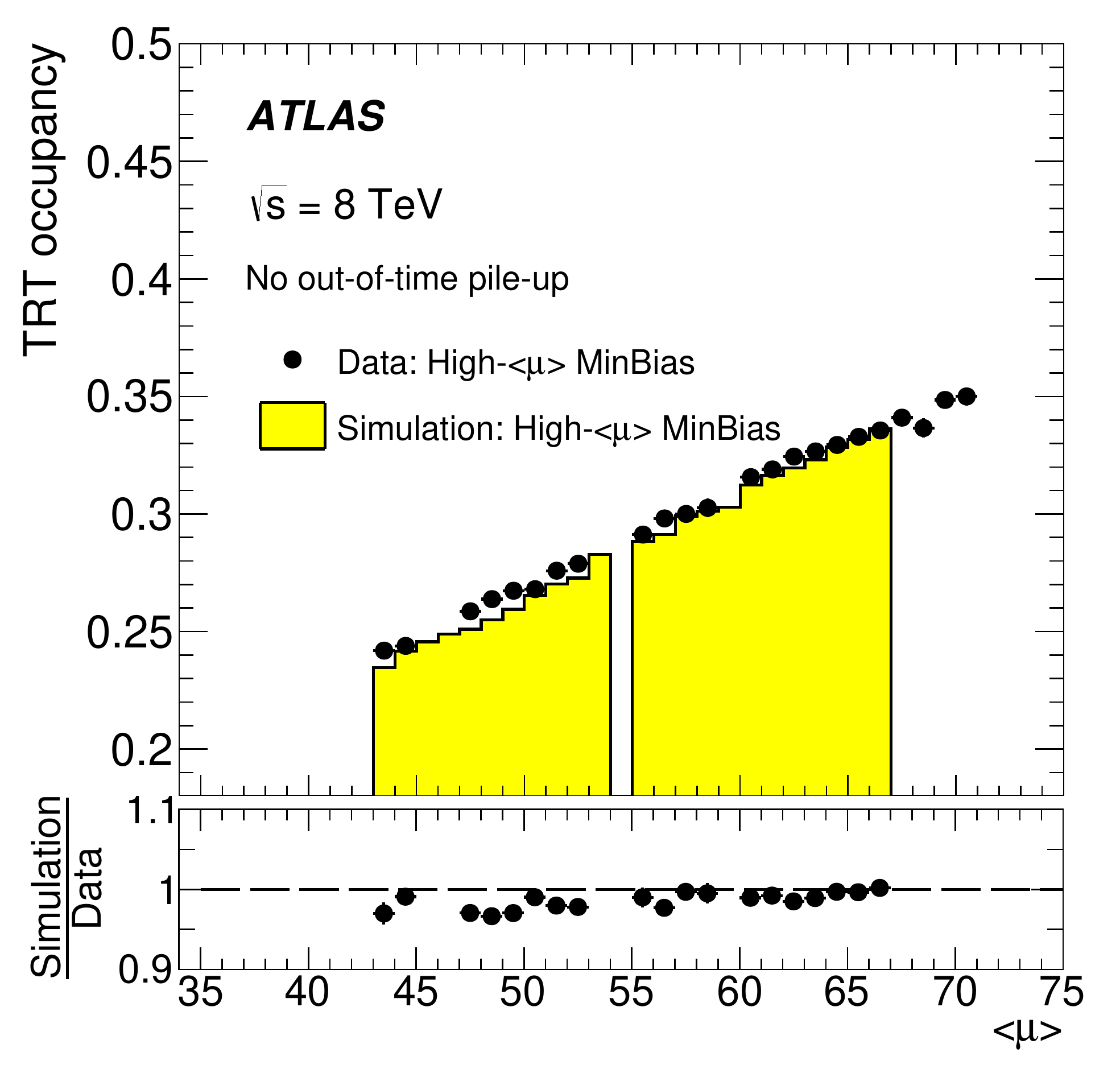}
 \caption{
  TRT straw occupancy as a function of $\langle\mu\rangle$ in the TRT from special high-$\langle\mu\rangle$ LHC fill
 (no pile-up from adjacent bunches) during a $pp$ data-taking period in 2012 operating with Xe-based gas mixture at $\sqrt{s} = 8~\TeV$, for data (solid circles) and simulation (filled histograms).}
  \label{fig:Occupancy_vs_mu_all}
 \end{figure}

Figure~\ref{fig:Occupancy_vs_mu_all} shows the correlation between the TRT occupancy and $\langle\mu\rangle$ for the high-$\langle\mu\rangle$ LHC fill. 
The simulation well describes the linear dependence of the TRT occupancy on the pile-up.
The average TRT occupancy for $\langle\mu\rangle = 70$ is about 35\% in these $pp$ collisions and can even reach a level of 50\% in some events due to the large fluctuation of the number of primary collisions. 

The centrality of nuclei collisions is generally used to characterise events in heavy ion runs. This parameter cannot be directly translated to 
the parameters used for $pp$ collisions, so the TRT occupancy is used as a reference to compare the results. In these collisions, TRT occupancies up to 90\% were obtained.

Even if exploring unprecedented occupancy conditions during Run 1, the results shown in this section still do not totally correspond to the design LHC running conditions: the high-$\langle\mu\rangle$ $pp$ LHC fill studied here was made of a single proton bunch in each direction colliding approximately every 25000\,ns, resulting in no out-of-time pile-up. 
Results are thus expected to show an optimistic view of TRT performance in high-occupancy conditions compared to the ones with larger out-of-time pile-up during LHC data-taking operation with a 25 ns bunch spacing. However they are still useful for understanding the performances in high in-time pile-up conditions.

In this section, two TRT tracking parameters (TRT track extension fraction and precision hit fraction) are studied using particle tracks with $p_{\mathrm{T}}$ in the $0.5$--$100~\GeV$ range.
The precision hit fraction is defined as the number of precision TRT hits (defined in Section~\ref{section:calib}) divided by the total number of hits on the selected track. 
In the standard ATLAS track-finding algorithm, tracks are first found in the silicon trackers (pixel and SCT) and then TRT hits are added to extend the number of measurements on the reconstructed track. 
If for some reason (for instance because of hard interactions of particles with the detector material) TRT hits are not found by the algorithm, then the track parameters are calculated using only the information from silicon detectors.
The TRT track extension fraction is the fraction of tracks reconstructed in the pixel and SCT detectors with their corresponding efficiencies~\cite{ATL-PHYS-PUB-2015-006}, which also have a continuation (extension) in the TRT. 
In these studies, at least 18 TRT hits are required to be on such tracks. 
The TRT track extension fraction reflects how efficiently the information from the TRT is used in the standard ATLAS track-finding algorithm.
For an extension to be accepted, several criteria have to be fulfilled, related to the minimum number of TRT hits, the fraction of precision hits on this extension, or the overall goodness of the fit after inclusion of the TRT segment.

\begin{figure}[!h]
\begin{center}
  \includegraphics[width=0.47\textwidth]{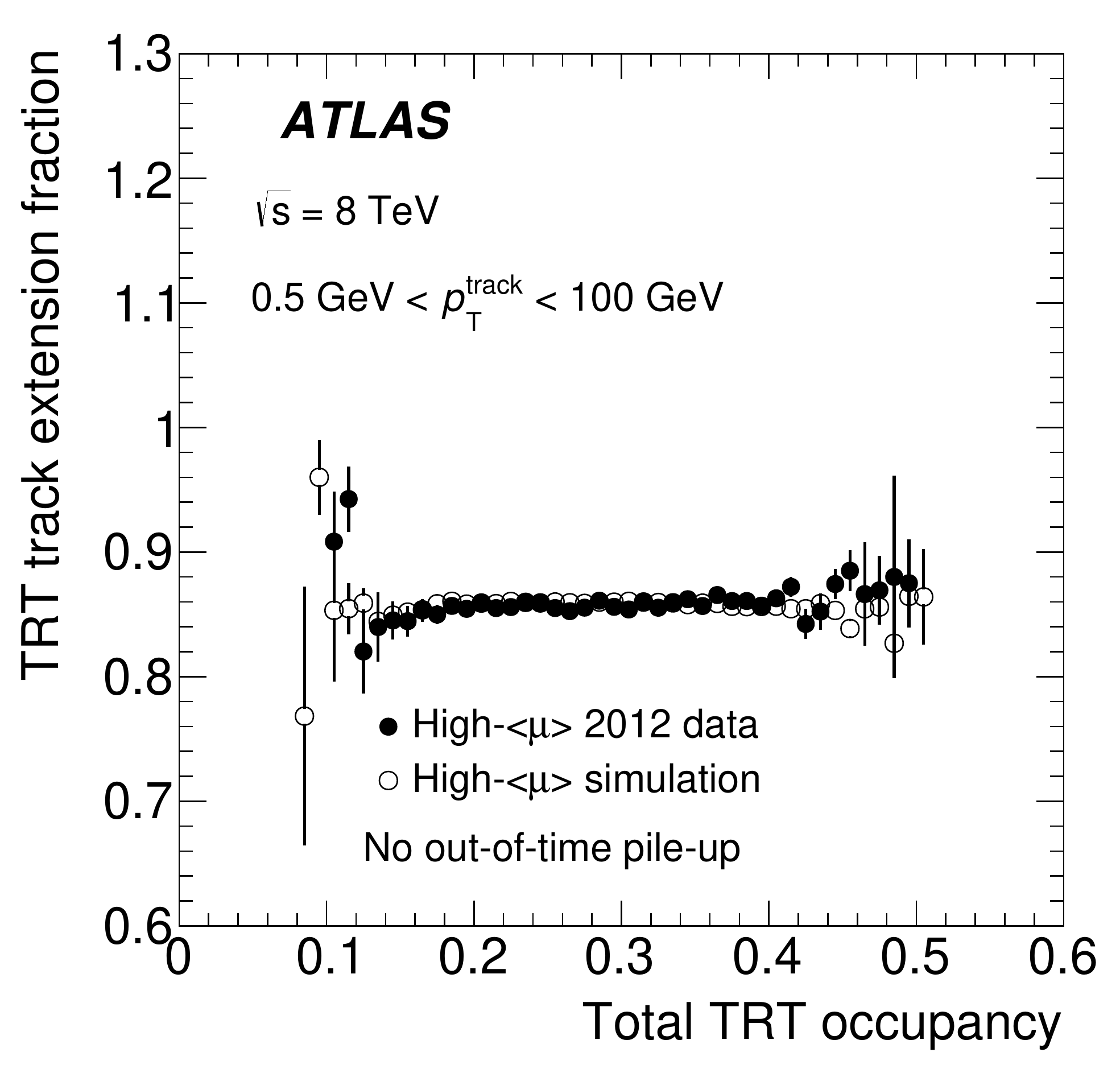}
  \label{fig:extension_fraction_vs_TRT_occupancy}
        \includegraphics[width=0.47\textwidth]{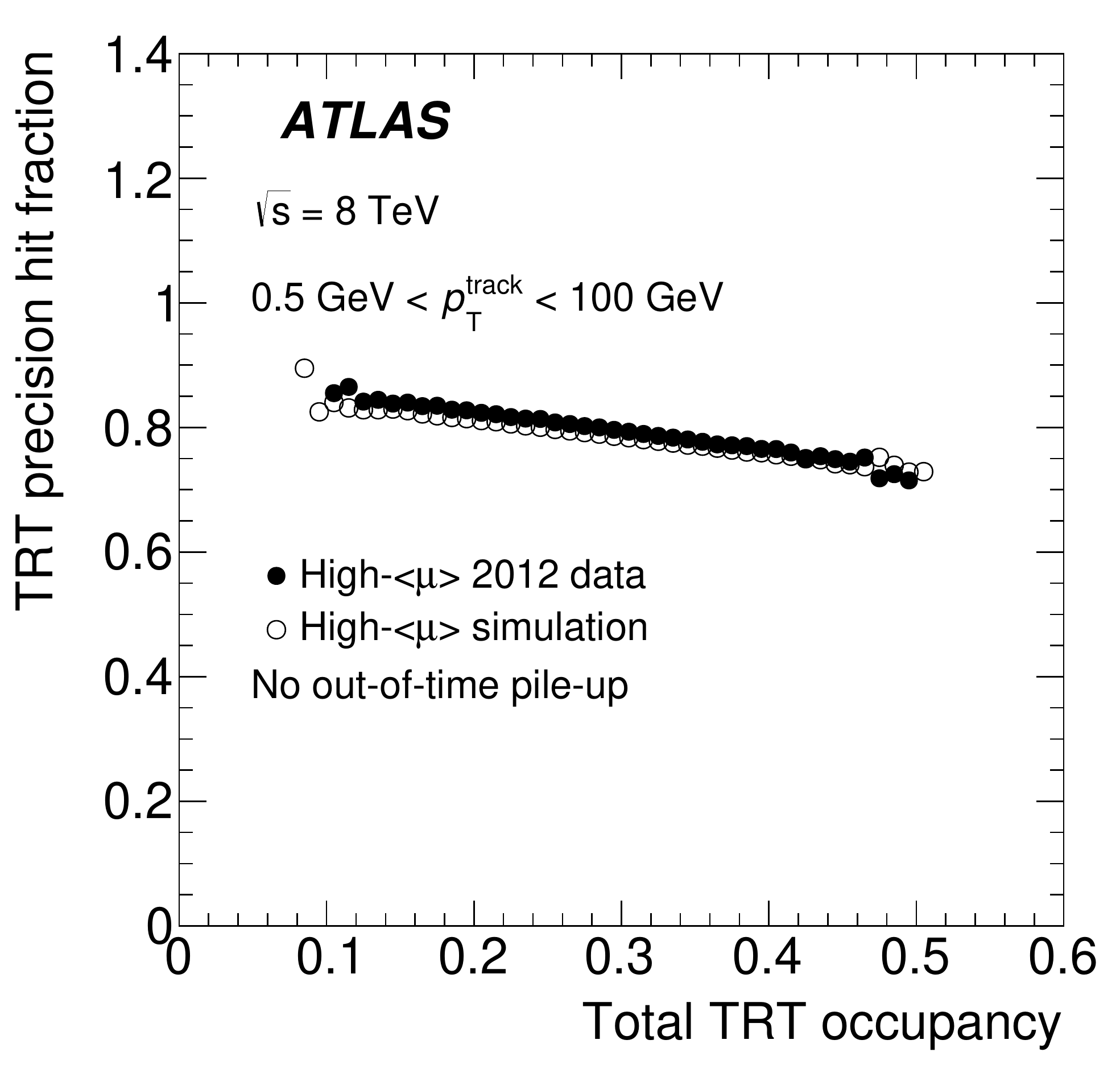}
        \label{fig:precision_hit_fraction_vs_TRT_occupancy}
\caption{
The TRT track extension fraction (left), and the precision hit fraction (right), as a function of the total TRT occupancy from minimum-bias events with $40 \le \langle\mu\rangle \le 70$ for $pp$ data (solid circles) and simulation (open circles) collected during the 2012 period operating with a Xe-based gas mixture. }
 \label{fig:ext_fr_prec_hit_fr_data_mc}
 \end{center}
 \end{figure}

Figure~\ref{fig:ext_fr_prec_hit_fr_data_mc} shows a comparison of data and simulation for $pp$ collisions for both the track extension fraction and precision hit fraction.
The track extension fraction is almost constant for occupancies up to 50\%. 
It happens that some tracks have a significant deviation from their ideal trajectory due to interactions with ID material. After these tracks are fitted, a lot of the TRT hits are lost because the real trajectory is not known well enough. These tracks generally fail to have the required minimum number of TRT hits on the extension. With increasing occupancy, additional hits are collected in the TRT track which are not from the particle of interest (background hits) but this causes the requirement to be fulfilled more frequently. This is why a small rise (2\%) of the TRT extension fraction is observed in this figure.

With further increase of the occupancy, the background hits start to shadow the genuine leading edges. This happens because the probability for a background track to cross the straw closer to the wire than the particle of interest increases and causes the leading edge to be earlier than the real one. However, these background hits give wrong drift circle measurements and thus cannot be considered as precision hits. The consequence is that the total number of precision hits on the track decreases with increasing occupancy, which leads to a reduction of the extension fraction due to the requirement on this quantity mentioned above. 
The two effects compensate for each other, producing a plateau in the extension fraction distribution. 
The decrease of the TRT precision hit fraction can be seen directly in the right panel of Figure~\ref{fig:ext_fr_prec_hit_fr_data_mc}. However, even for a detector with about 50\% straw occupancy, more than 75\% of the track hits correspond to precision hits.
The simulation describes the data well.

 \begin{figure}[!h]
    \begin{center}
   \includegraphics[width=0.48\textwidth]{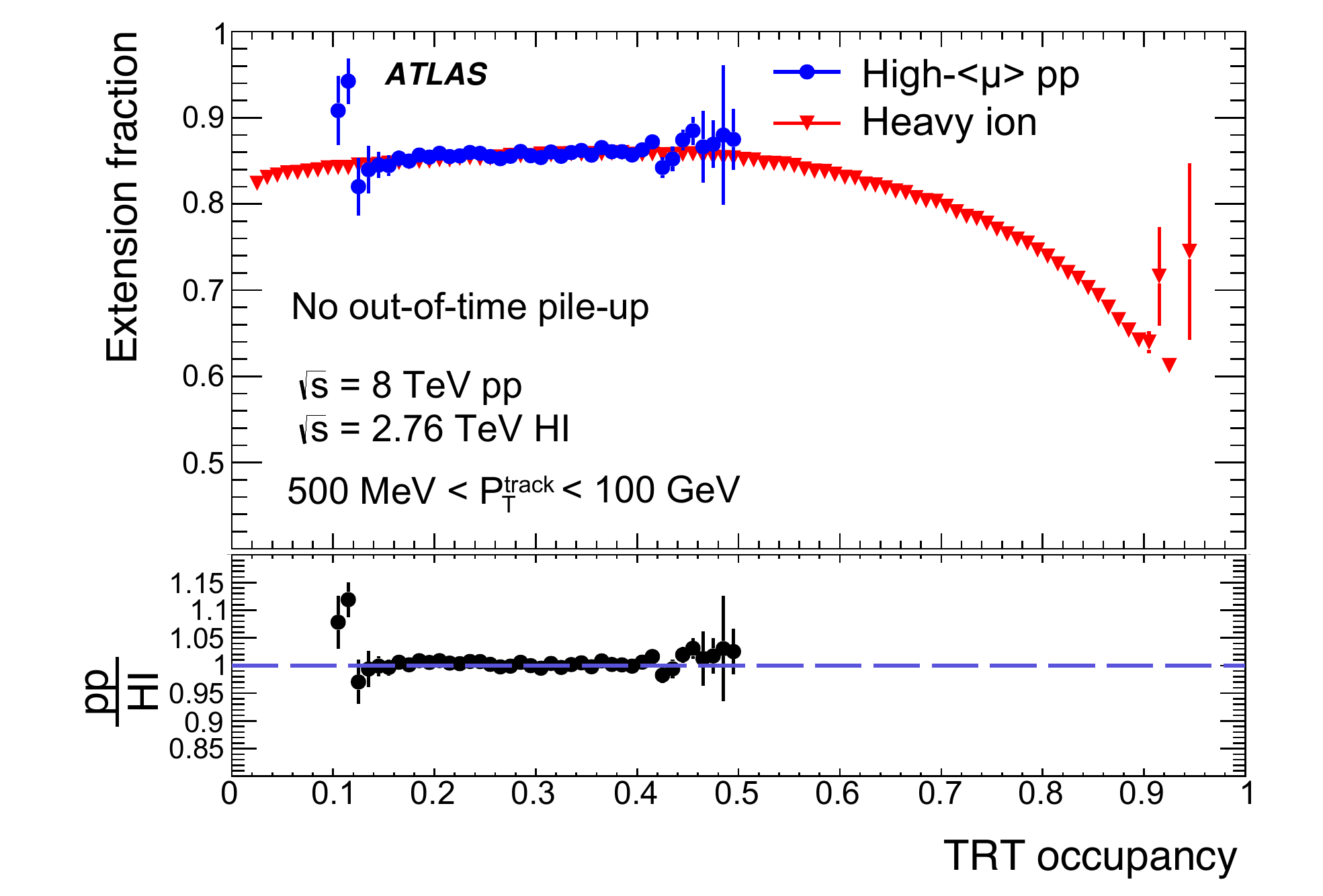}
   \label{fig:extension fraction}
         \includegraphics[width=0.48\textwidth]{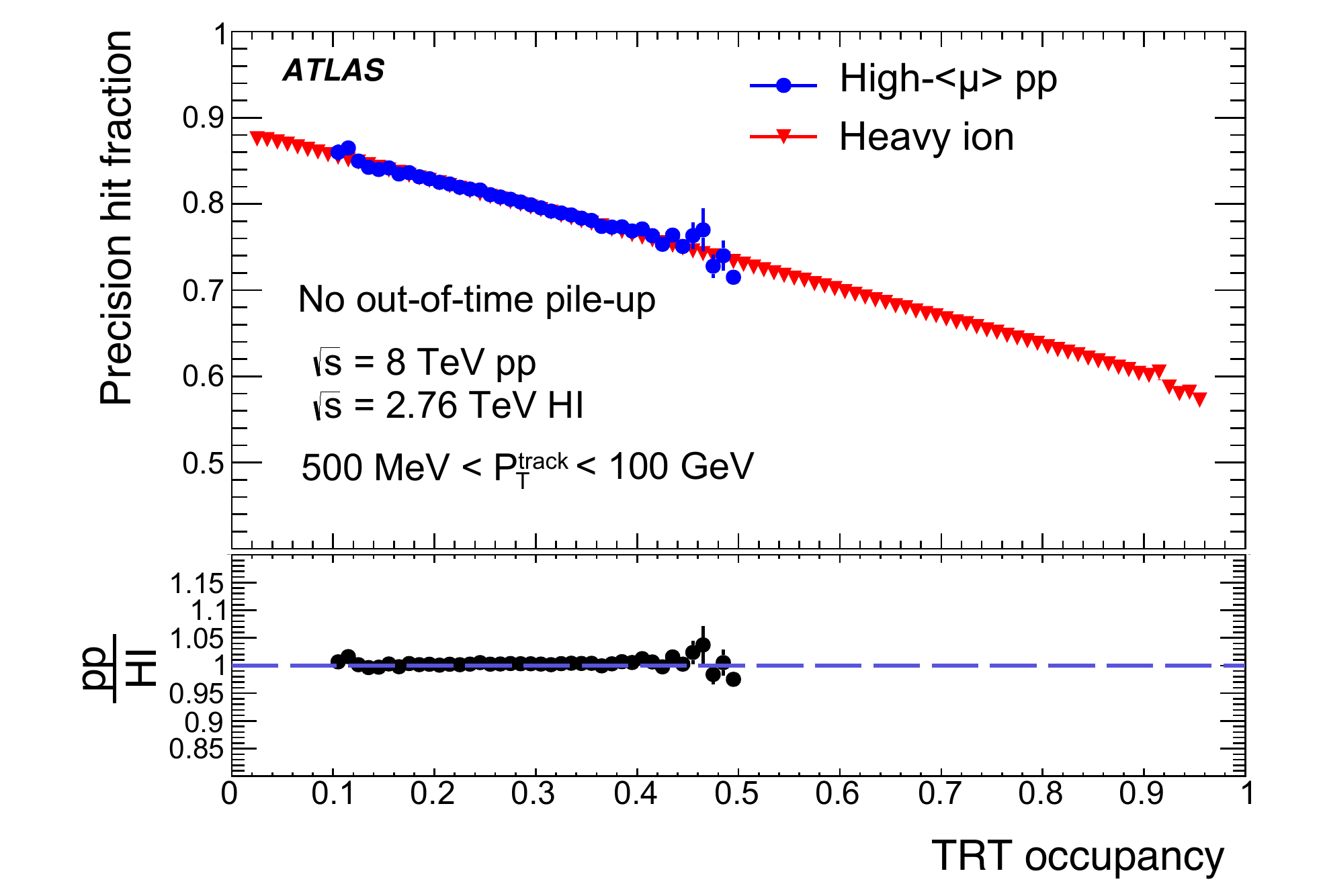}
         \label{fig:precision hit fraction} \\
  \centering
  \includegraphics[width=0.48\columnwidth]{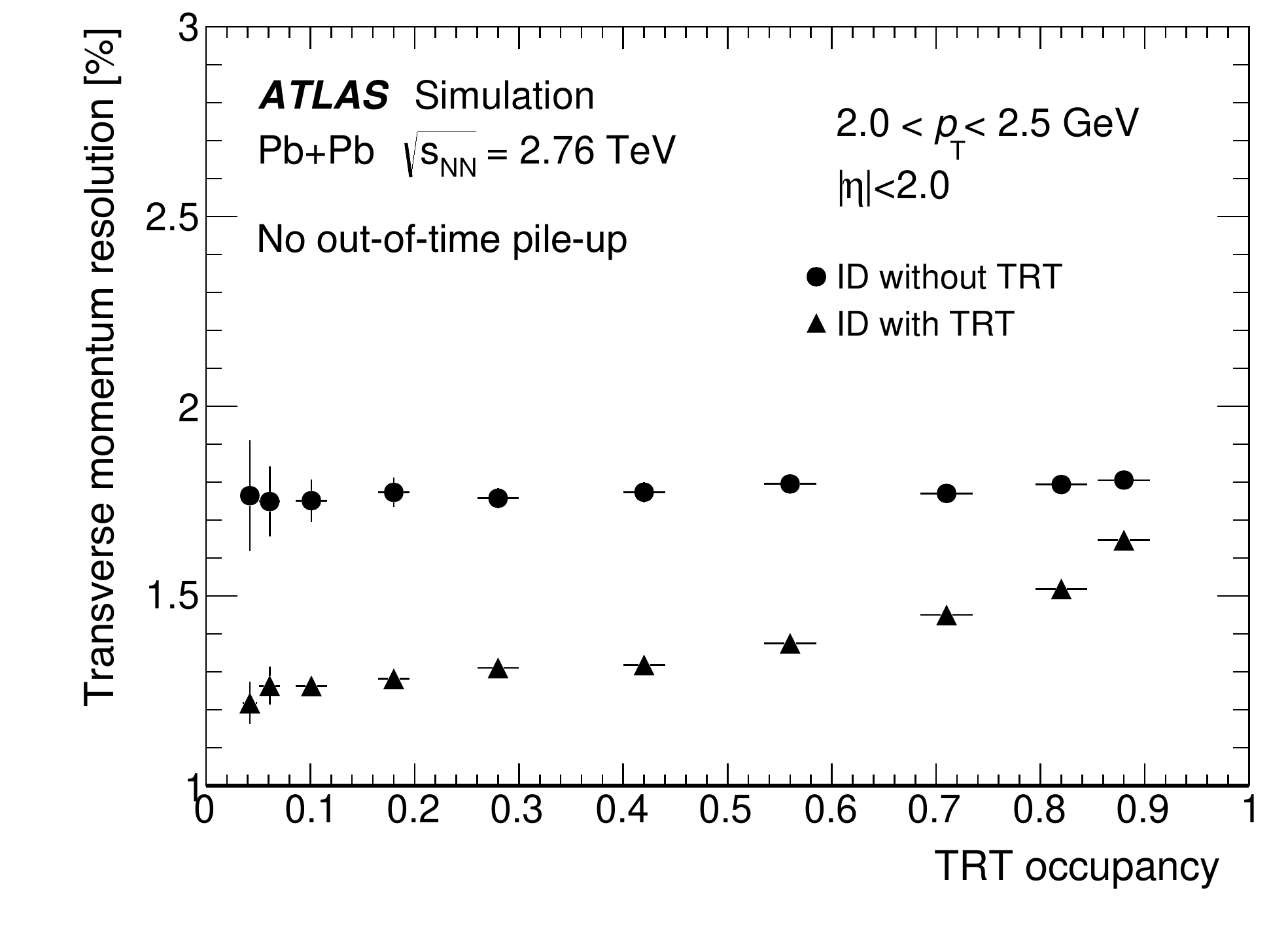}
  \label{fig:pt_resolution}
          \caption{The TRT track extension fraction (top left), and the precision hit fraction (top right),
        as a function of the total TRT occupancy from minimum-bias events with $40 \le \langle\mu\rangle \le 70$ for $pp$ and heavy-ion collision events
during the 2011--2012 operation with a Xe-based gas mixture. The $\pT$ resolution of ID tracks as a function of the TRT occupancy (bottom)
with (triangle marker) and without (circle marker) TRT measurements for heavy-ion simulations with a Xe-based gas mixture. Tracks with TRT measurements 
correspond to 70\% of the tracks in the ID in such high-occupancy conditions as shown in the top left panel.}
   \label{fig:extension fraction and precision hit fraction}
  \end{center}
\end{figure}

Figure~\ref{fig:extension fraction and precision hit fraction} shows the track extension fraction and precision hit fraction as a 
function of the TRT occupancy for proton--proton and heavy-ion collisions.
The required fraction of precision hits on the extension is 30\% in heavy-ion collisions while it is 50\% in $pp$ collisions.
Good agreement between the two running conditions is seen in the overlapping range of occupancies for both the extension fraction and precision hit fraction distributions. 
 
The TRT contribution to the particle momentum reconstruction accuracy is studied in heavy-ion simulations. 
The true particle momentum, $p_\mathrm{T}^\mathrm{true}$, is compared with the reconstructed momentum, $p_{\mathrm {T}}^{\mathrm{reco}}$, for the particle momentum 
range of $2<p_{\mathrm{T}}<2.5~\GeV$. This low $p_{T}$ range is chosen because this is where the occupancy effects are expected to 
be the most pronounced in this particular simulation sample. The $\pT$ resolution is defined as the width $\sigma$ of a Gaussian function fitted to 
the core part (within $\pm 1.5\sigma$) of the distribution of $R_{q/\pT}$ 
in Eq.(\ref{eq:equation_pt_reso}) and is shown in the bottom panel of Figure~\ref{fig:extension fraction and precision hit fraction}. 

\begin{equation}
  \label{eq:equation_pt_reso}
R_{q/\pT} = \left( \frac{q}{p_{\mathrm{T}}^{\mathrm{reco}}} - \frac{q}{p_{\mathrm{T}}^{\mathrm{true}}} \right) / \frac{q}{p_{\mathrm{T}}^{\mathrm{true}}}
\end{equation}

The case where ID tracks are reconstructed without including TRT measurements is compared to the case where the tracks are required to have at 
least 20 TRT hits of which at least 30\% are precision hits.
Including the TRT hits in the particle momentum measurement improves the momentum resolution by about 25\%. 
The contribution of the TRT is found to be almost constant for occupancies up to 50\%.
For occupancies larger than 60\%, there is some reduction of the TRT contribution to the track measurement accuracy, but even at 90\% occupancy 
the TRT improves the ID momentum resolution by about 10\% for tracks with extensions in the TRT, which applies to about 70\% of the tracks in 
such high-occupancy conditions in heavy-ion collisions.

\section{Performance in dense track environments}
\label{section:jets}
One of the most challenging tasks for a tracking detector is finding and reconstructing tracks 
in regions with high track density, as in energetic jet cores where tracks can be very close to each other. 
The high pile-up LHC conditions during Run 1 and Run 2 lead to high TRT hit occupancies, making this task particularly difficult, when the distance between tracks in the jet cores becomes comparable to the straw diameter. Studies of the tracking inside jets is presented in this section, using minimum bias events produced during high-$\langle\mu\rangle$ $pp$ LHC fills in 2012. Events with at least one high-energy jet ($30 < \pt < 400~\GeV$),
and with at least one primary vertex with two associated tracks are selected.

Each track is required to satisfy a number of selection criteria: $|\eta|$ has to be less than 2.0 (excluding $0.625 < |\eta| < 1.07$ corresponding to the TRT barrel to end-cap transition regions), its transverse momentum has to be larger than $2~\GeV$ and it needs to have hits in all active pixel layers and at least five hits in the SCT. 
For the track extension in the TRT, more than 18 TRT hits on the track are required. 

\begin{figure}[!ht]
  \centering
  \includegraphics[width=0.42\columnwidth]{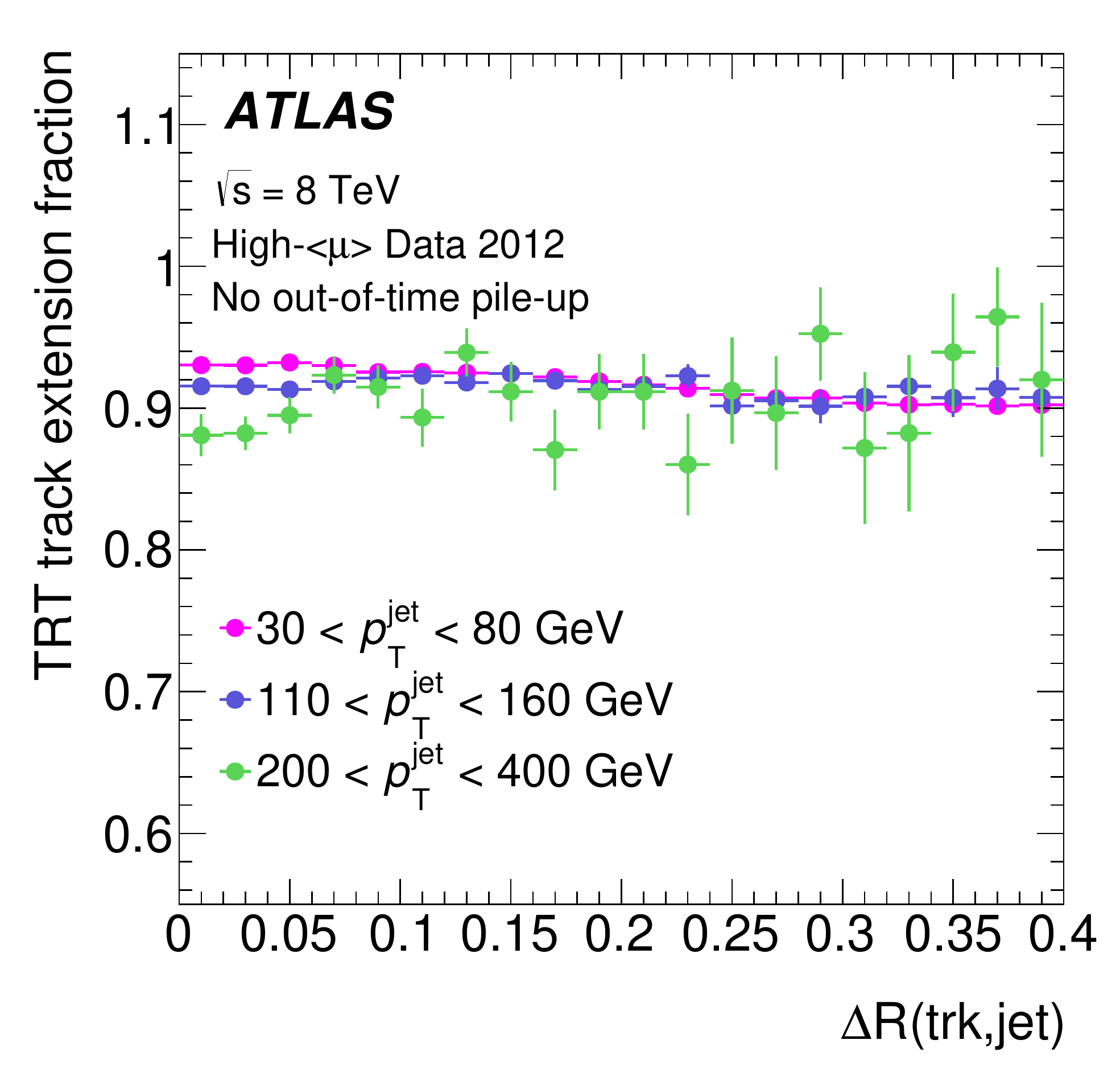}
  \includegraphics[width=0.42\columnwidth]{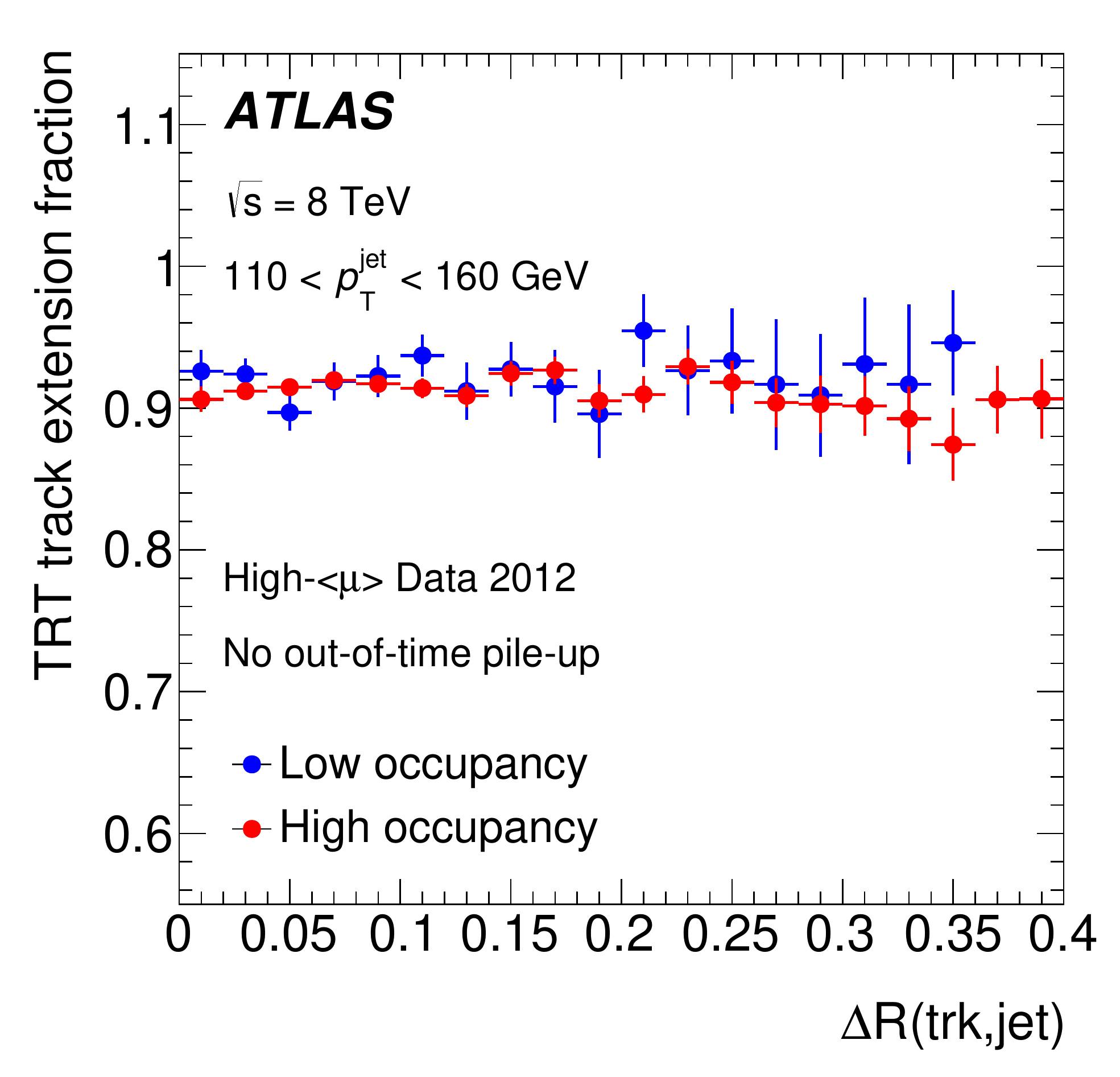}
  \caption{TRT track extension fraction as a function of $\Delta R$ in jet cores for three jet $\pt$ ranges in data (left). TRT track extension fraction as a function of $\Delta R$ in jet cores ( $110 < p_{\mathrm{T}}^{\mathrm{jet}} < 160~\GeV$) for low (less than 24\%) and high-occupancy (34\% to 54\%) conditions in data (right). Data events were collected during the 2012 period operating with a Xe-based gas mixture.}
 \label{fig:extension_fraction_vs_dR-ref}
 \end{figure}

Jets are reconstructed from topological calorimeter clusters~\cite{Aad:2016upy} using the anti-$k_t$ algorithm~\cite{Cacciari:2008gp} 
with a radius parameter of $R=0.4$ and their energies are corrected for the non-compensating response of the calorimeter and the effects of inactive material using energy- and 
$\eta$-dependent calibration factors, based on collision data and simulation~\cite{Aad:2011he}. 
The distance of closest approach of each track to the primary vertex, located on the nominal beam line, is required to be less than 1.5 mm in 
both the transverse plane and along the $z$ direction. Tracks are assigned to jets using a geometrical algorithm. 
If the distance in $\eta$--$\phi$ space between the track and the jet, $\Delta R\text{(trk, jet)}=\sqrt{\Delta \eta^{2}+\Delta \phi^{2}}$, 
is less than the radius parameter used in the jet reconstruction ($R = 0.4$), the tracks are assigned to the jet.

The jet core becomes denser with increasing momentum, making the track finding more challenging.
In order to study this effect, data are studied  as a function of $\Delta R\text{(trk, jet)}$ and for three jet $\pt$ regions: $30 < \pt < 80~\GeV$, 
$110 < \pt < 160~\GeV$ and $200 < \pt < 400~\GeV$. 

\begin{figure}[!ht]
  \centering
  \includegraphics[width=0.42\columnwidth]{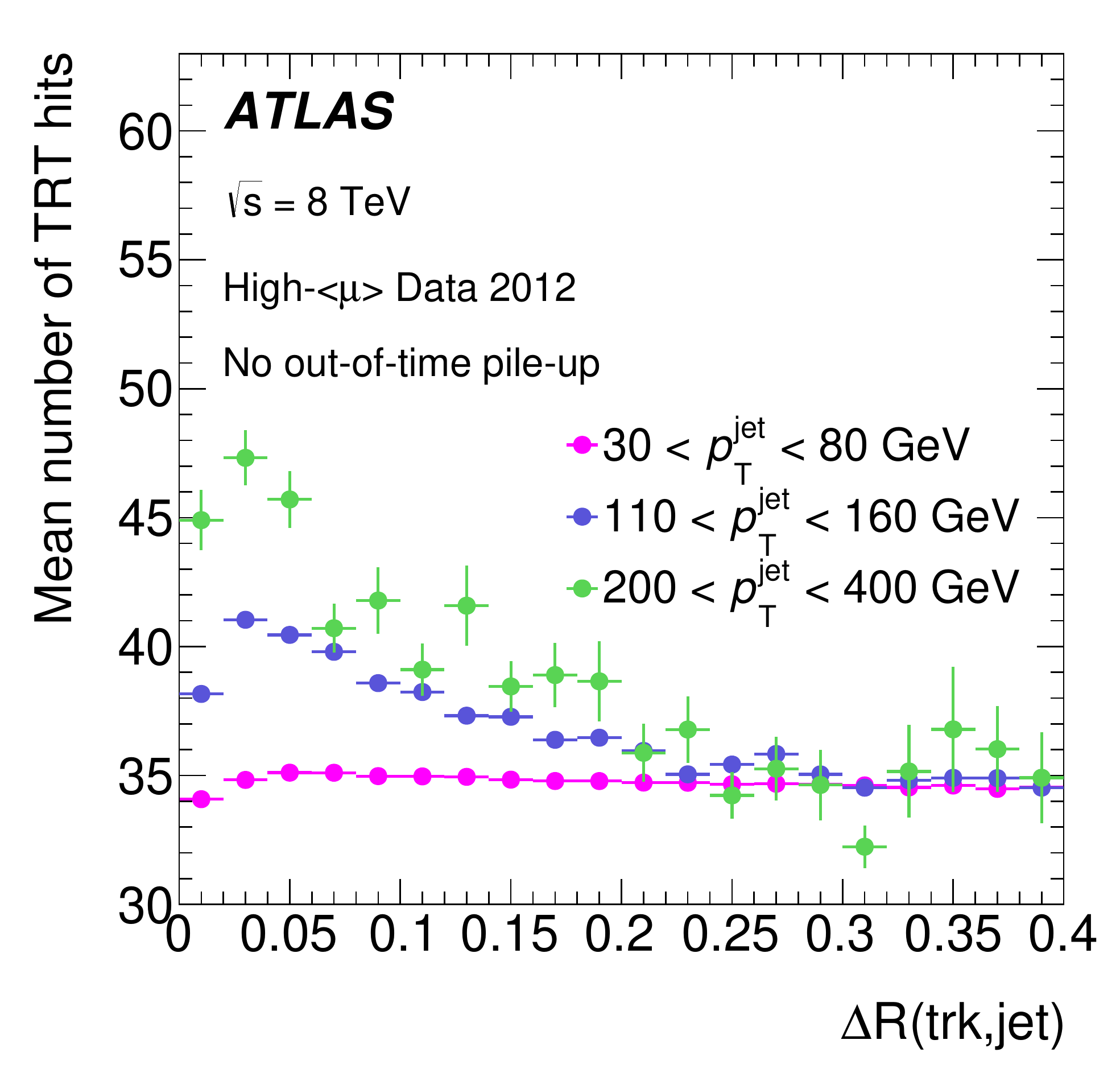}
  \caption{Total averaged number of TRT hits on tracks found in a given bin of $\Delta R$ = 0.02 as a function of $\Delta R$ in jet cores for three jet $\pt$ ranges for data events collected during the 2012 period operating with a Xe-based gas mixture.}
  \label{fig:nTRThits_vs_dR-ref}
 \end{figure}

Figure~\ref{fig:extension_fraction_vs_dR-ref}
shows the TRT track extension fraction as a function of $\Delta R$ in jet cores, averaged over the entire  $\langle\mu\rangle$ range ($40 \le \langle\mu\rangle \le 70$).
The TRT track extension fraction is found to be almost constant in the cores of energetic jets.
Figure~\ref{fig:extension_fraction_vs_dR-ref} also shows the TRT track extension fraction for low (less than 24\%) and high TRT occupancy ranges (between 34\% and 54\%).
It is observed that higher TRT occupancies do not affect the track extension fraction in jet cores. 

\begin{figure}[!ht]
  \centering
  \includegraphics[width=0.42\columnwidth]{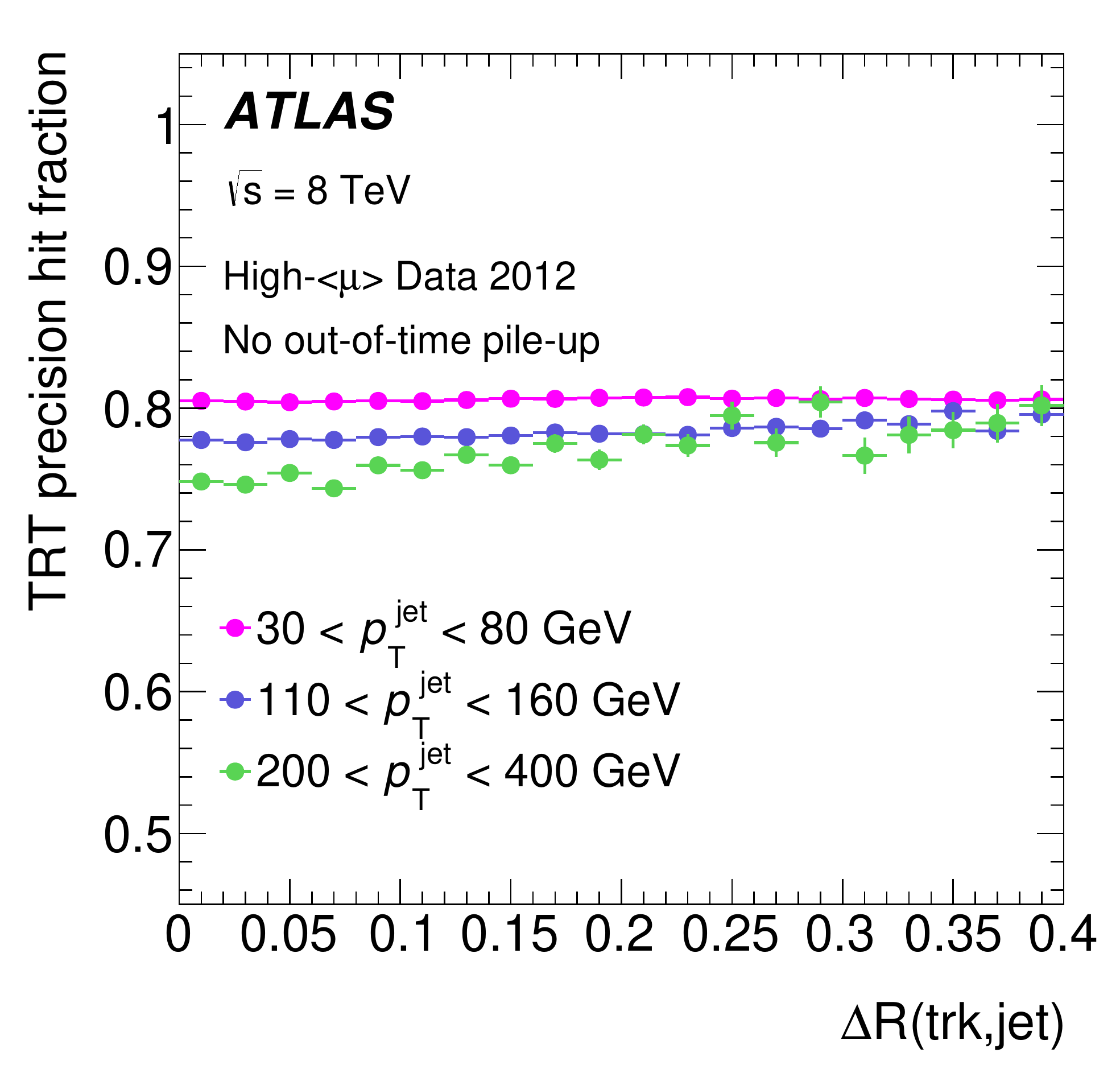}
  \caption{TRT precision hit fraction on reconstructed tracks as a function of $\Delta R$ in jet cores for three jet $\pt$ ranges and for data events collected during the 2012 period operating with a Xe-based gas mixture.}
  \label{fig:precision_hit_fraction_vs_dR}
 \end{figure}

\begin{figure}[!ht]
   \begin{center}
            \includegraphics[width=0.4\columnwidth]{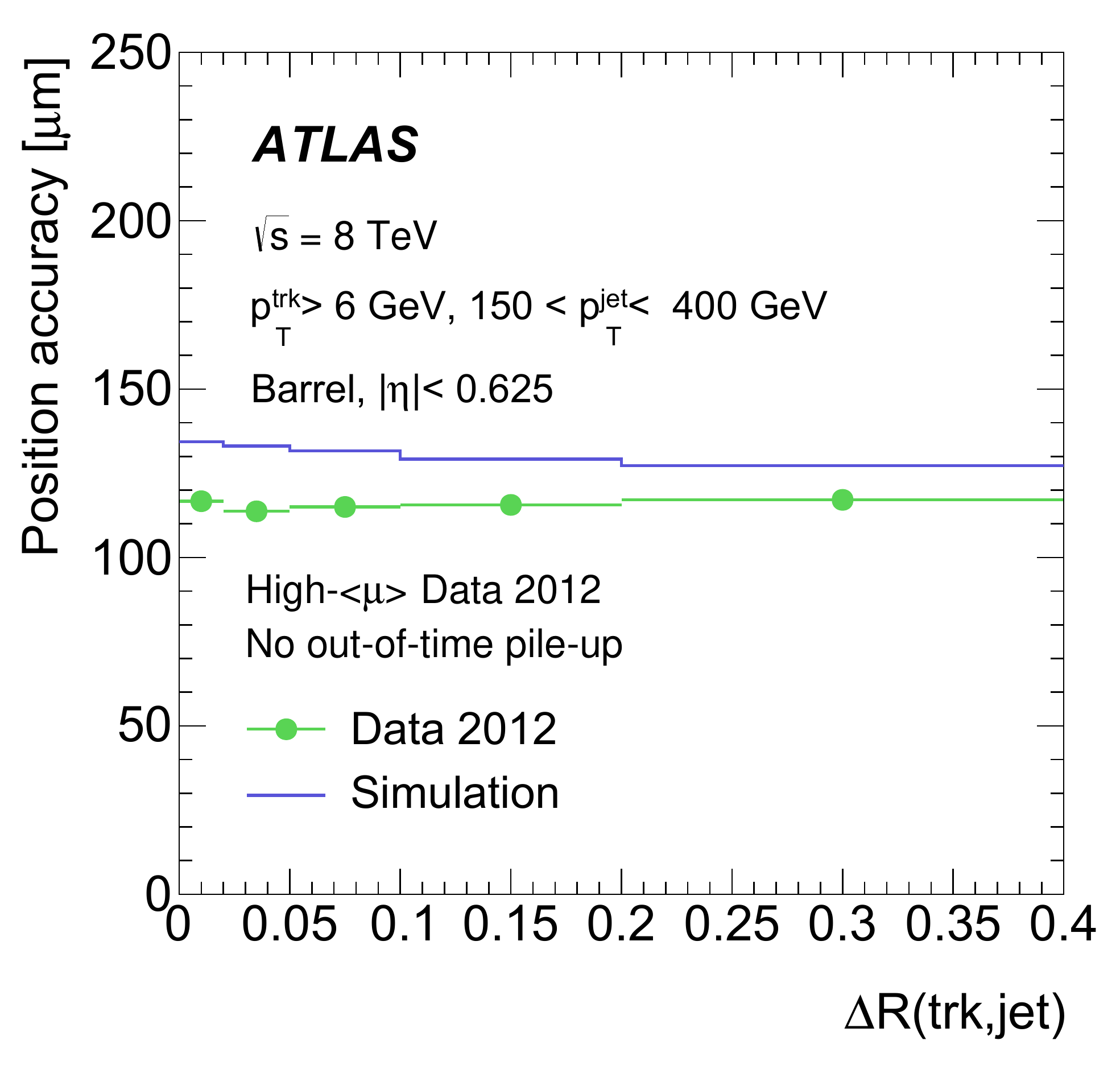}
            \label{fig:track_position_accuracy_vs_dR_barrel}
  \includegraphics[width=0.4\columnwidth]{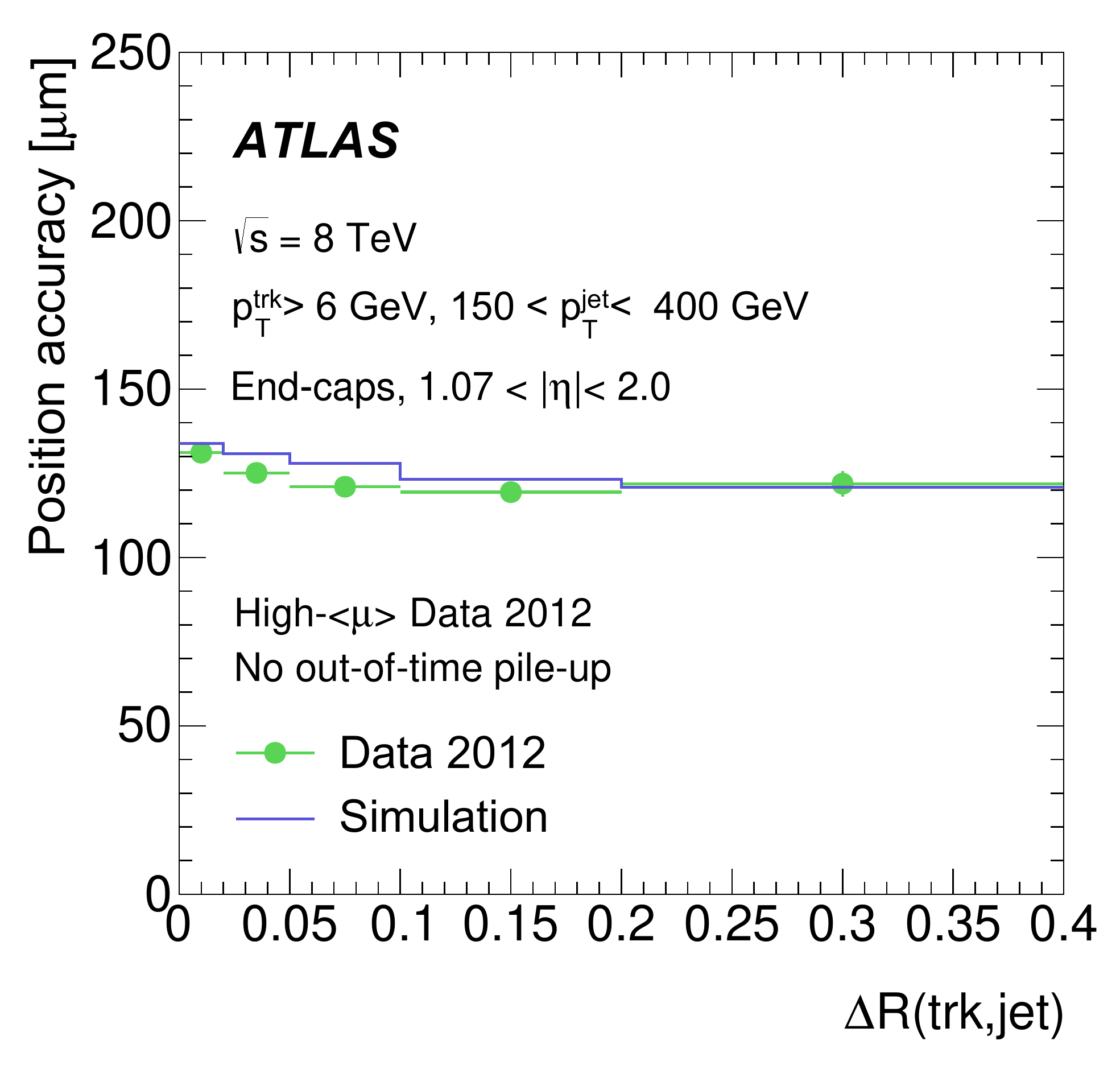}
   \label{fig:track_position_accuracy_vs_dR_endcaps}
  \caption{The straw track position measurement accuracy as a function of $\Delta R$ for tracks with $\pT > 6~\GeV$ for the barrel (left) and end-caps (right), for simulation (histograms) and data (points) events collected during the 2012 period operating with a Xe-based gas mixture. See the text for details of the discrepancies between data and simulation.}
  \label{fig:track_position_accuracy_vs_dR}
   \end{center}
 \end{figure}

In order to characterise the conditions inside the jets, the total averaged number of TRT hits belonging to reconstructed tracks within each $\Delta R = 0.02$ 
interval around the jet core was studied.
Figure~\ref{fig:nTRThits_vs_dR-ref} shows the result for different jet transverse momentum ranges and
for the entire $\langle\mu\rangle$ range in data.
Far from the jet core, the average number of hits is about 35 and corresponds to only one reconstructed track.
The observed increase in the total number of TRT hits closer to the jet core is related to the corresponding increase in track density, leading to a greater probability to have more than one track reconstructed within $\Delta R = 0.02$. 
Also, higher jet energies correspond to a larger number of tracks in the jet core, also resulting in a greater total number of hits in the TRT.

Figure~\ref{fig:precision_hit_fraction_vs_dR}
shows the TRT precision hit fraction as a function of $\Delta R$ in jet cores averaged over the entire $\langle\mu\rangle$ range for data events.
Even in the core of the most energetic jets in this sample, where the average number of tracks is large, the precision 
hit fraction is similar to the one outside of the jet core. This implies that the reconstruction algorithm is able to find the correct hits and properly reconstruct the tracks even when tracks are close to each other.

In order to study the straw track position measurement accuracy in dense environments, jets with $150 < \pT < 400~\GeV$
in the high-$\langle\mu\rangle$ run were selected and tracks were required to have $\pT > 6~\GeV$. 
The results for different $\Delta R$ bins are shown in Figure~\ref{fig:track_position_accuracy_vs_dR} for the TRT end-caps and barrel.
The plot shows that the straw track position measurement accuracy for high-$\pT$ tracks is approximately constant
even within the dense cores of the most energetic jets in this sample. 
Similarly to Figure~\ref{fig:hit_res_eta_unbiased_mu_5_10-ref}, the simulation predicts slightly higher values (about 10\%) for the straw track position measurement accuracy in the core of jets for the barrel, for the reasons mentioned in Section \ref{section:xenon}. 

\section{Conclusion}
\label{section:conclusion}

The TRT tracking performance at low and high occupancies based on proton-proton, heavy ion and proton-lead collisions during Run 1 of the LHC operation is presented. The TRT operation with either a xenon-based or an argon-based gas mixture is studied and similar tracking performance is found for the two gas mixtures. For both gas mixtures, the straw efficiency is found to be larger than 96\% and the straw track position measurement accuracy in the straws better than 130 
$\mu \mathrm{m}$ in low occupancy conditions. 
Straw efficiency and measurement accuracy are studied as a function of $\eta$, $\langle\mu\rangle$ and occupancy for the Xe-based gas mixture 
and compared to simulations, and a fair level of agreement is observed everywhere. For the Ar-based gas mixture, the dependence of the straw efficiency and the straw track position measurement accuracy on the LL threshold are studied. For threshold variations between 55 eV and 100 eV in data, the straw efficiency 
changes only by 1--2\% and the straw track position measurement accuracy varies from $\sim$110 $\mu$m to 130 $\mu$m, respectively. 
No significant degradation of these parameters is found up to occupancies of about 20\% ($\langle\mu\rangle$ up to 30) for 50 ns LHC bunch spacing. 

For higher occupancies the TRT tracking performance parameters are studied using special $pp$ and heavy ion data-taking fills with 
respectively 25000 ns and 200 ns bunch spacing, resulting in no out-of-time pile-up in events. 
The number of precision hits on track is reduced from 85\% to 75\% when the TRT occupancy increases from 10\% to 50\%.
This is expected due to the shadowing effect from pile-up particles. However, the TRT significantly contributes to the particle momentum measurement by the ATLAS ID up 
to straw occupancies of 90\%, which are reached in heavy ion collisions.

Additional studies of the track characteristics in jet cores as a function of the jet $p_{\mathrm{T}}$ and the track distance $\Delta R\text{(trk, jet)}$ from the jet axis are performed. 
It is shown that the reconstruction algorithm is able to properly reconstruct the tracks inside high-energy jet cores.

\section*{Acknowledgements}


We thank CERN for the very successful operation of the LHC, as well as the
support staff from our institutions without whom ATLAS could not be
operated efficiently.

We acknowledge the support of ANPCyT, Argentina; YerPhI, Armenia; ARC, Australia; BMWFW and FWF, Austria; ANAS, Azerbaijan; SSTC, Belarus; CNPq and FAPESP, Brazil; NSERC, NRC and CFI, Canada; CERN; CONICYT, Chile; CAS, MOST and NSFC, China; COLCIENCIAS, Colombia; MSMT CR, MPO CR and VSC CR, Czech Republic; DNRF and DNSRC, Denmark; IN2P3-CNRS, CEA-DSM/IRFU, France; GNSF, Georgia; BMBF, HGF, and MPG, Germany; GSRT, Greece; RGC, Hong Kong SAR, China; ISF, I-CORE and Benoziyo Center, Israel; INFN, Italy; MEXT and JSPS, Japan; CNRST, Morocco; FOM and NWO, Netherlands; RCN, Norway; MNiSW and NCN, Poland; FCT, Portugal; MNE/IFA, Romania; MES of Russia and NRC KI, Russian Federation; JINR; MESTD, Serbia; MSSR, Slovakia; ARRS and MIZ\v{S}, Slovenia; DST/NRF, South Africa; MINECO, Spain; SRC and Wallenberg Foundation, Sweden; SERI, SNSF and Cantons of Bern and Geneva, Switzerland; MOST, Taiwan; TAEK, Turkey; STFC, United Kingdom; DOE and NSF, United States of America. In addition, individual groups and members have received support from BCKDF, the Canada Council, CANARIE, CRC, Compute Canada, FQRNT, and the Ontario Innovation Trust, Canada; EPLANET, ERC, ERDF, FP7, Horizon 2020 and Marie Sk{\l}odowska-Curie Actions, European Union; Investissements d'Avenir Labex and Idex, ANR, R{\'e}gion Auvergne and Fondation Partager le Savoir, France; DFG and AvH Foundation, Germany; Herakleitos, Thales and Aristeia programmes co-financed by EU-ESF and the Greek NSRF; BSF, GIF and Minerva, Israel; BRF, Norway; CERCA Programme Generalitat de Catalunya, Generalitat Valenciana, Spain; the Royal Society and Leverhulme Trust, United Kingdom.

The crucial computing support from all WLCG partners is acknowledged gratefully, in particular from CERN, the ATLAS Tier-1 facilities at TRIUMF (Canada), NDGF (Denmark, Norway, Sweden), CC-IN2P3 (France), KIT/GridKA (Germany), INFN-CNAF (Italy), NL-T1 (Netherlands), PIC (Spain), ASGC (Taiwan), RAL (UK) and BNL (USA), the Tier-2 facilities worldwide and large non-WLCG resource providers. Major contributors of computing resources are listed in Ref.~\cite{ATL-GEN-PUB-2016-002}.

\printbibliography

\newpage 

\begin{flushleft}
{\Large The ATLAS Collaboration}

\bigskip

M.~Aaboud$^\textrm{\scriptsize 137d}$,
G.~Aad$^\textrm{\scriptsize 88}$,
B.~Abbott$^\textrm{\scriptsize 115}$,
J.~Abdallah$^\textrm{\scriptsize 8}$,
O.~Abdinov$^\textrm{\scriptsize 12}$,
B.~Abeloos$^\textrm{\scriptsize 119}$,
S.H.~Abidi$^\textrm{\scriptsize 161}$,
O.S.~AbouZeid$^\textrm{\scriptsize 139}$,
N.L.~Abraham$^\textrm{\scriptsize 151}$,
H.~Abramowicz$^\textrm{\scriptsize 155}$
% ...
\bigskip
\\
% ...
$^{8}$ Department of Physics, The University of Texas at Arlington, Arlington TX, United States of America\\
% ...
$^{12}$ Institute of Physics, Azerbaijan Academy of Sciences, Baku, Azerbaijan\\
% ...
$^{88}$ CPPM, Aix-Marseille Universit{\'e} and CNRS/IN2P3, Marseille, France\\
% ...
$^{115}$ Homer L. Dodge Department of Physics and Astronomy, University of Oklahoma, Norman OK, United States of America\\
% ...
$^{119}$ LAL, Univ. Paris-Sud, CNRS/IN2P3, Universit{\'e} Paris-Saclay, Orsay, France\\
% ...
$^{137}$ $^{(a)}$ Facult{\'e} des Sciences Ain Chock, R{\'e}seau Universitaire de Physique des Hautes Energies - Universit{\'e} Hassan II, Casablanca; $^{(b)}$ Centre National de l'Energie des Sciences Techniques Nucleaires, Rabat; $^{(c)}$ Facult{\'e} des Sciences Semlalia, Universit{\'e} Cadi Ayyad, LPHEA-Marrakech; $^{(d)}$ Facult{\'e} des Sciences, Universit{\'e} Mohamed Premier and LPTPM, Oujda; $^{(e)}$ Facult{\'e} des sciences, Universit{\'e} Mohammed V, Rabat, Morocco\\
% ...
$^{139}$ Santa Cruz Institute for Particle Physics, University of California Santa Cruz, Santa Cruz CA, United States of America\\
% ...
$^{151}$ Department of Physics and Astronomy, University of Sussex, Brighton, United Kingdom\\
% ...
$^{155}$ Raymond and Beverly Sackler School of Physics and Astronomy, Tel Aviv University, Tel Aviv, Israel\\
% ...
$^{161}$ Department of Physics, University of Toronto, Toronto ON, Canada\\
% ...
$^{a}$ Also at Department of Physics, King's College London, London, United Kingdom\\
$^{b}$ Also at Institute of Physics, Azerbaijan Academy of Sciences, Baku, Azerbaijan\\
$^{c}$ Also at Novosibirsk State University, Novosibirsk, Russia\\
$^{d}$ Also at TRIUMF, Vancouver BC, Canada\\
$^{e}$ Also at Department of Physics {\&} Astronomy, University of Louisville, Louisville, KY, United States of America\\
% ...
\end{flushleft}


\end{document}